\documentclass{svmult}

\usepackage{graphicx,amsmath,latexsym,amsfonts}

\font\sqi=cmssq8
\def\DR{\rm I\kern-1.45pt\rm R}
\def\DC{\kern2pt {\hbox{\sqi I}}\kern-4.2pt\rm C}
\def\DH{\rm I\kern-1.5pt\rm H\kern-1.5pt\rm I}

\newcommand{\be}{\begin{equation}}
\newcommand{\ee}{\end{equation}}
\newcommand{\bea}{\begin{eqnarray}}
\newcommand{\eea}{\end{eqnarray}}

\begin{document}

\title*{A Journey Through Garden Algebras}
\author{Stefano~Bellucci\inst{1}, Sylvester~James~Gates\inst{2}, Jr.,
Emanuele~Orazi\inst{1,3}}

\authorrunning{S.~Bellucci, S.~J.~Gates,~Jr., ~E.~Orazi }

\institute{INFN-Laboratori Nazionali di Frascati, \\Via E. Fermi
40, C.P. 13, 00044 Frascati, Italy\\
\texttt{bellucci@lnf.infn.it},~\texttt{orazi@lnf.infn.it}
\vspace{0.5cm} \and University of Maryland,
Physics Department, \\
Rm. 4125, College Park, MD 20742-4111\\
\texttt{gatess@wam.umd.edu} \vspace{0.5cm}
\and Universit\`a di Roma ``Tor Vergata", Dipartimento di Fisica, ,\\
Via della Ricerca Scientifica 1, 00133 Roma, Italy } \maketitle

\vspace{2.5cm}

\begin{abstract}
The main purpose of these lectures is to give a pedagogical
overview on the possibility to classify and relate off-shell
linear supermultiplets in the context of supersymmetric mechanics.
A special emphasis is given to a recent graphical technique that
turns out to be particularly effective for describing many aspects of
supersymmetric mechanics in a direct and simplifying way.
\end{abstract}

\pagebreak

\setcounter{equation}{0}
\section*{Introduction}

Sometimes problems in mathematical physics go unresolved for long
periods of time in mature topics of investigation. During this
World Year of Physics which commemorates the pioneering efforts of
Albert Einstein, it is perhaps appropriate to note the
irreconciliability of the symmetry group of Maxwell Equations with
that of Newton's Equation (via his second law of motion) was one
such problem.  The resolution of this problem, of course, led to
one of the greatest revolutions in physics.  This piece of history
suggests a lesson on what can be the importance of problems that
large numbers of physicists regard as unimportant or unsolvable.

\vspace{0.5cm}

 In light of this episode, the presentation which
follows hereafter is focused on a problem in supersymmetry that
has long gone unresolved and seems generally regarded as one of
little importance.  While there is no claim or pretension that
this problem has the importance of the one resolved by the
brilliant genius of Einstein,  it is a problem that perhaps holds
the key to a more mathematically complete understanding of the
area known as ``supersymmetry.''

\vspace{0.5cm}

The topic of supersymmetry is over thirty years old now. It has
been vigorously researched by both mathematicians and physicists.
During this entire time, this subject has been insinuated into a
continuously widening array of increasingly sophisticated
mathematical models.  At the end of this  stream of development
lies the mysterious topic known as ``M-theory.''  Accordingly, it
may be thought that all fundamental issue regarding this area have
already a satisfactory resolution.

\vspace{0.5cm}

However, as surprising as it may seem, in fact very little is
known about the representation theory of supersymmetry required
for the classification of irreducible superfield theories in a
manner that allows for quantization consistent with a manifest
realization of supersymmetry.

\vspace{0.5cm}

Superspace is to supersymmetry as Minkowski space is to the
Lorentz group.  Superspace provides the most natural geometrical
setting in which to describe supersymmetrical theories.  Almost no
physicist would utilize the component of Lorentz four-vectors or
higher rank tensor to describe relativistic physics.  Yet, the
analog of this is common practice in describing supersymmetrical
theory.  This is so because `component fields' are the predominant
language by which most discussions of supersymmetry are couched.

\vspace{0.5cm}

One fact that hides this situation is that much of the language
used to describe supersymmetrical theories appears to utilize the
superspace formalism.  However, this appearance is deceiving. Most
often what appears to be a superspace presentation is actually a
component presentation in disguise.  A true superspace formulation
of a theory is one that uses `unconstrained' superfields as their
fundamental variables.  This is true of an tiny subset of the
discussions of supersymmetrical theories and is true of none of
the most interesting such theories involving superstrings.

\vspace{0.5cm}

This has led us to the belief that
possibly some important fundamental issues regarding supersymmetry
have yet to be properly understood.  This belief has been the
cause of periodic efforts that have returned to this issue.
Within the last decade this investigation has pointed toward two
new tools as possibly providing a fresh point of departure for the
continued study (and hopefully ultimate resolution) of this
problem.    One of these tools has relied on a totally new setting
in which to understand the meaning of supersymmetry.  This has led
to the idea that the still unknown complete understanding of the
representation theory of supersymmetry lies at the intersection of
the study of Clifford algebras and K-theory.  In particular, a
certain class of Clifford algebras (to which the moniker ${\cal
{GR}} ({\rm d}, \,N)$ have been attached) provides a key to making
such a connection.  Within the confines of an interdisciplinary
working group that has been discussing these problems, the term
``garden algebra'' has been applied
 to the symbolic name
${\cal {GR}} ({\rm d}, \,N)$.  It has also been shown that these
Clifford algebras natural lead to a graphical representation
somewhat akin to the root and weight spaces seen in the
classification of compact Lie algebras.  These graphs have been
given the name ``Adinkras.''  The topic of this paper will be
introducing these new tools for the study of supersymmetry
representation theory.

\vspace{0.5cm}

\setcounter{equation}{0}

\section{${\cal GR}(d, \, N)$ Algebras}

\subsection{Geometrical interpretation of ${\cal GR}(d, \, N)$ algebras}

In a field theory, boson and fermions are to be regarded as
diffeomorphisms generating two different vector spaces; the
supersymmetry generators are nothing but sets of linear maps
between these spaces. Following this picture we can include a
supersymmetric theory in a more general geometrical framework
defining the collection of diffeomorphisms \be \phi_i : R
\rightarrow R^{d_L} ,\quad i~=~1,..,d_L \ee \be \psi_{\hat \alpha}
: R \rightarrow R^{d_R} ,\quad i~=~1,..,d_R \ee where the one
dimensional dependence reminds us that we restrict our attention
to mechanics. The free vector spaces generated by
$\{\phi_i\}_{i~=~1}^{d_L}$ and $\{\psi_{\hat \alpha}\}_{{\hat
\alpha}~=~1}^{d_R}$ are respectively ${\cal V}_L$ and ${\cal
V}_R$, isomorphic to $R^{d_L}$ and $R^{d_R}$. For matrix
representations in the following, the two integers are restricted
to the case ${d_L}$ $=$ ${d_R}$ $=$ ${d}$. Four different linear
mappings can act on ${\cal V}_L$ and ${\cal V}_R$ \bea &{\cal
M}_L:{\cal V}_{L} \rightarrow {\cal V}_{R} ,\quad
{\cal M}_R :{\cal V}_{R} \rightarrow {\cal V}_{L} \nonumber \\
&{\cal U}_L :{\cal V}_{L} \rightarrow {\cal V}_{L} ,\quad
{\cal U}_R :{\cal V}_{R} \rightarrow {\cal V}_{R}
\eea
with linear maps space dimensions
\bea
&dim{\cal M}_L ~=~dim{\cal M}_R ~=~d_{R} d_{L}~=~d^{2}~, \nonumber \\
&dim~{\cal U}_L ~=~{d_{L}}^2~=~d^{2}~,\quad dim~{\cal U}_R
~=~{d_{R}}^2~=~d^{2}~, \eea as a consequence of linearity. In
order to relate this construction to a general real $(\equiv {\cal
GR})$ algebraic structure of dimension $d$ and rank $N$ denoted by
${\cal GR}({\rm d}, N)$, two more requirements need to be added.
\begin{enumerate}
\item Let us define the generators of ${\cal GR}({\rm d}, N)$ as
the family of $N+N$ linear maps \footnote{Notice that in previous
works on the subject \cite{jim2}\cite{jim0}, the maps $L_I$ and
$R_K$ where exchanged, so that $L_I\in\{{\cal M}_R\}$ and
$R_K\in\{{\cal M}_L\}$} \bea
&L_I\in\{{\cal M}_L\},\quad I~=~1,..,N \nonumber \\
&R_K\in\{{\cal M}_R\},\quad K~=~1,..,N \eea such that for all
$I,K~=~1,..,N$ we have \bea
&L_I\circ R_K+L_K\circ R_I~=~-2\delta_{IK}I_{{\cal V}_R}~, \nonumber \\
&R_I\circ L_K+R_K\circ L_I~=~-2\delta_{IK}I_{{\cal V}_L}~,
\label{cond1} \eea where $I_{{\cal V}_L}$ and $I_{{\cal V}_R}$ are
identity maps on ${\cal V}_L$ and ${\cal V}_R$. Equations
(\ref{cond1})will later be embedded into a Clifford algebra but
one point has to be emphasized, we are working with real objects.
\newline $~~$ \newline



\item After equipping ${\cal V}_L$ and ${{\cal V}_R}$ with euclidean inner products $\langle \cdot,\cdot \rangle_{{\cal V}_L}$ and $\langle \cdot,\cdot \rangle_{ {\cal V}_R}$ respectively, the generators satisfy the property

\be \langle \phi, R_{I} (\psi) \rangle_{{\cal V}_L}~=~-\langle
L_{I} (\phi) , \psi \rangle_{{\cal V}_R} ~,~~~~ \forall
(\phi,\psi)\in {\cal V}_L \oplus {\cal V}_R ~.\label{cond2} \ee
This condition relates $L_I$ to the hermitian conjugate of $R_I$,
namely ${R_I}^{\dagger}$, defined as usual by

\be \langle \phi, R_{I} (\psi) \rangle_{{\cal V}_L}~=~\langle
R_{I}^{\dagger} (\phi) , \psi \rangle_{{\cal V}_R} \ee so that \be
R_{I}^{\dagger}~=~R_{I}^t~=~-L_{I}~. \ee
\end{enumerate}
The role of $\{{\cal U}_{L}\}$ and $\{{\cal U}_{R}\}$ maps is to
connect different representations once a set of generators defined
by conditions (\ref{cond1}) and (\ref{cond2}) has been chosen.
Notice that $\left( R_{I}L_{J}\right)_{i}^{~j}\in{{\cal U}_{L}}$
and $\left(
L_{I}R_{J}\right)_{\hat{\alpha}}^{~\hat{\beta}}\in{{\cal U}_{R}}$.
Let us consider ${\cal A}\in{\{{\cal U}_{L}\}}$ and ${\cal
B}\in{\{{\cal U}_{R}\}}$ such that

\bea
&{\cal A}:\phi \rightarrow \phi'~=~{\cal A}\phi \nonumber \\
&{\cal B}:\psi \rightarrow \psi'~=~{\cal B}\psi
\eea
then, taking the ${\cal V}_L$ sector as example, we have
\bea
&\langle \phi, R_{I} (\psi) \rangle_{{\cal V}_L} \rightarrow \langle {\cal A}\phi, R_{I} {\cal B}(\psi) \rangle_{{\cal V}_L}= \nonumber \\
&=\langle \phi,{\cal A}^{{\dagger}} R_{I} {\cal B}(\psi) \rangle_{{\cal V}_L}= \nonumber \\
&=\langle \phi, R_{I}' (\psi) \rangle_{{\cal V}_L} \eea so a
change of representation transforms the generators in the
following manner: \bea
&~~L_{I} \rightarrow L_{I}'~=~{\cal B}^{{\dagger}}L_{I}{\cal A} \nonumber \\
&R_{I} \rightarrow R_{I}'~=~{\cal A}^{{\dagger}}R_{I}{\cal B}~.
\eea In general, equations (\ref{cond1}) and (\ref{cond2}) do not
identify a unique set of generators. Thus, an equivalence relation
has to be defined on the space of possible sets of generators, say
$\{L_{I},R_{I}\}\sim \{L_{I}',R_{I}'\}$ if and only if there exist
${\cal A}\in\{{\cal U}_{L}\}$ and ${\cal B}\in\{{\cal U}_{R}\}$
such that $L'~=~{\cal B}^{{\dagger}}L_{I}{\cal {A}}$ and
$R'~=~{\cal A}^{{\dagger}}R_{I}{\cal {B}}$.

\vspace{0.5cm}

Now we want to show how a supersymmetric theory arises. Algebraic
derivations are defined by \bea
&\delta_{\epsilon}\phi_{i}~=~i\epsilon^I{(R_I)}^{~\hat{\alpha}}_{i}\psi_{\hat{\alpha}} \nonumber \\
&\delta_{\epsilon}\psi_{\hat{\alpha}}~=~-\epsilon^I{(L_I)}_{\hat{\alpha}}^{~i}\partial_{\tau}\phi_{i}
\label{algder} \eea where the real valued fields
$\{\phi_i\}_{i~=~1}^{d_L}$ and $\{\psi_{\hat \alpha}\}_{{\hat
\alpha}~=~1}^{d_R}$ can be interpreted as bosonic and fermionic
respectively. The fermionic nature attributed to the ${\cal
V}_{R}$ elements implies that ${\cal M}_{L}$ and ${\cal M}_{R}$
generators, together with supersymmetry transformation parameters
$\epsilon^{I}$, anticommute among themselves. Introducing the
$d_{L}+d_{R}$ dimensional space ${\cal V}_{L}\oplus{V}_{R}$ with
vectors

\be
\Psi~=~\left(
\begin{array}{c}
\phi\\ \psi
\end{array}
\right)~, \label{rep}
\ee
equation (\ref{algder}) reads

\be
\delta_{\epsilon}(\Psi)~=~
\left(
\begin{array}{c}
i\epsilon R\psi\\
\epsilon L{\partial}_{\tau}\phi
\end{array}
\right)
\ee
so that
\be
\left[\delta_{\epsilon_{1}},\delta_{\epsilon_{2}}\right]\Psi~=~
i\epsilon_{1}^{I}\epsilon_{2}^{J}
\left(
\begin{array}{c}
R_{I}L_{J}\partial_{\tau}\phi\\
L_{I}R_{J}\partial_{\tau}\psi
\end{array}
\right)
-i\epsilon_{2}^{J}\epsilon_{1}^{I}
\left(
\begin{array}{c}
R_{J}L_{I}\partial_{\tau}\phi\\
L_{J}R_{I}\partial_{\tau}\psi
\end{array}
\right)~=~-2i\epsilon_{1}^{I}\epsilon_{2}^{I}\partial_{\tau}\Psi~,
\label{commsusy} \ee utilizing that we have classical
anticommuting parameters and that equations (\ref{cond1}) hold. It
is important to stress that components of (\ref{rep}) can be
interpreted as superfield components, so it is as if we were
working with a particular superfield multiplet containing only
these physical bosons and fermions. From (\ref{commsusy}) it is
clear that $\delta_{\epsilon}$ acts as a supersymmetry generator,
so that we can set

\be
\delta_{Q}\Psi~:=~\delta_{\epsilon}\Psi~=~i\epsilon^{I}Q_{I}\Psi
\ee which is equivalent to writing \bea
&\delta_{Q}\phi_{i}~=~i\left(\epsilon^{I}Q_{I}\psi\right)_{i}~, \nonumber \\
&\delta_{Q}\psi_{\hat{\alpha}}=~i\left(\epsilon^{I}Q_{I}\phi\right)_{\hat{\alpha}}~,
\eea with

\be
Q_{I}~=~\left(
\begin{array}{cc}
0& R_{I}\\ L_{I}H& 0
\end{array}
\right)~, \ee where $H=i\partial_{\tau}$. As a consequence of
(\ref{commsusy}) a familiar anticommutation relation appears

\be \{Q_{I},Q_{J}\}~=~-2i\delta_{IJ}H~, \ee confirming that we are
talking about genuine supersymmetry. Once the supersymmetry is
recognized, we can associate to the algebraic derivations
(\ref{algder}) the variations defining the scalar supermultiplets.
However, the choice (\ref{algder}) is not unique: one can check
that \bea
\delta_{Q}\xi_{\hat{\alpha}}~&=&~\epsilon^I{(L_I)}_{\hat{\alpha}}^{~i}F_{i}~, \nonumber \\
\delta_{Q}F_{i}~&=&~-i\epsilon^I{(R_I)}^{~\hat{\alpha}}_{i}\partial_{\tau}\xi_{\hat{\alpha}}~,
\label{spinmult} \eea is another proposal linked to ordinary
supersymmetry as the previous one. In this case we will refer to
the supermultiplet defined by (\ref{spinmult}) as the spinorial
one.

\subsection{Twisted representations}

The construction outlined above suffers from an ambiguity in the
definition of superfield components $\left(\phi_{i},
\psi_{\hat{\alpha}}\right)$ and $\left(\xi_{\hat{\alpha}},
A_{i}\right)$ due to the possibility of exchanging the role of $R$
and $L$ generators, giving rise to the new superfields
$\left(\phi_{\hat{\alpha}}, \psi_{i}\right)$ and $\left(\xi_{i},
A_{\hat{\alpha}}\right)$ with the same supersymmetric properties
of the previous ones. The variations associated to these twisted
versions are, respectively \bea
\delta_{Q}\phi_{\hat{\alpha}}~&=&~i\epsilon^I{(L_I)}_{\hat{\alpha}}^{~i}\psi_{i} \nonumber \\
\delta_{Q}\psi_{i}~&=&~-\epsilon^I{(R_I)}^{~\hat{\alpha}}_{i}\partial_{\tau}\phi_{\hat{\alpha}}
\label{scaltwist} ~,\eea and \bea
\delta_{Q}\xi_{i}~&=&~\epsilon^I{(R_I)}^{~\hat{\alpha}}_{i}F_{\hat{\alpha}}~, \nonumber \\
\delta_{Q}F_{\hat{\alpha}}~&=&~-i\epsilon^I{(L_I)}_{\hat{\alpha}}^{~i}\partial_{\tau}\xi_{i}
~.\label{spintwist}\eea The examples above mentioned are just some
cases of a wider class of inequivalent representations, referred
to as ``twisted" ones. The possibility to pass from a
supermultiplet to its twisted version is realized by the so called
``mirror maps". Moreover, it is possible to define superfields in
a completely different manner by parameterizing the supermultiplet
using component fields which take value in the algebra vector
space. We will refer to these objects as Clifford algebraic
superfields. An easy way to construct this kind of representations
is tensoring the superspace $\{{\cal V}_L\}\oplus~\{{\cal V}_R\}$
with $\{{\cal V}_L\}$ or $\{{\cal V}_R\}$. For instance, if we
multiply from the right by $\{{\cal V}_L\}$ then we have \be
(\{{\cal V}_L\}\oplus~\{{\cal V}_R\})\otimes~\{{\cal
V}_L\}=\{{\cal U}_L\}\oplus~\{{\cal M}_L\}\ee whose fields content
is \bea
&\phi_{i}^{~j}\in \{~{\cal U}_L\}~,\nonumber\\
&\psi_{\hat{\alpha}}^{~i}\in \{~{\cal M}_L\}~, \label{cliffsupfield}
\eea with supersymmetry transformations \bea
&\delta_{Q}\phi_{i}^{~j}~=~-i\epsilon^I{(R_I)}_{i}^{~\hat{\alpha}}\psi_{\hat{\alpha}}^{~j}~, \nonumber \\
&\delta_{Q}\psi_{\hat{\alpha}}^{~i}~=~\epsilon^I{(L_I)}_{\hat{\alpha}}^{~j}\partial_{\tau}\phi_{j}^{~i}~,
\label{CAtransf} \eea still defining a scalar supermultiplet. An
analogous structure can be assigned to $\{~{\cal
U}_R\}\oplus\{~{\cal M}_L\}$, $\{~{\cal U}_L\}\oplus\{~{\cal
M}_R\}$ and $\{~{\cal U}_R\}\oplus\{~{\cal M}_R\}$ type
superspaces. Even in these cases, twisted versions can be
constructed applying considerations similar to those stated above.
The important difference between the Clifford algebraic
superfields approach and the ${\cal V}_R\oplus{\cal V}_L$
superspace one, resides in the fact that in the latter case the
number of bosonic fields (which actually describe coordinates)
increases with the number of supersymmetric charges, while in the
first case there is a way to make this not happen, allowing for a
description of arbitrary extended supersymmetric spinning particle
systems, as it will be shown in the third section.

\subsection{${\cal GR}(d, \, N)$  algebras representation theory}

It is time to clarify the link with real Clifford $\Gamma$-matrices of Weyl type ($\equiv$ block skew diagonal) space which is easily seen to be

\be
\Gamma_{I}~=~\left(
\begin{array}{cc}
0& R_{I}\\ L_{I}& 0
\end{array}
\right)~. \label{gammamatr} \ee In fact, due to (\ref{cond1}),
$\Gamma$-matrices in (\ref{gammamatr}) satisfy

\be \{\Gamma_{I},\Gamma_{J}\}~=~-2i\delta_{IJ}I~,~~~~\forall
I,J~=~1,..,N~, \label{cliffalg1} \ee which is the definition of
Clifford algebras. One further $\Gamma$-matrix, namely

\be
\Gamma_{N+1}~=~\left(
\begin{array}{cc}
I& 0\\ 0& -I
\end{array}
\right),\label{GN+1} \ee can be added. Therefore the complete
algebra obeys the relationships

\be
\{\Gamma_{A},\Gamma_{B}\}~=~-2i\eta_{AB}I~,~~~~\forall A,B~=~1,..,N+1~, \label{cliffalg2}
\ee
where

\be
\eta_{AB}~=~diag(\underbrace{1,..,1}_{N},-1 )~. \label{eta}
\ee In the following we assume that $A,B$ indices run from $1$ to
$N+1$ while $I,J$ run from $1$ to $N$. The generator (\ref{GN+1}),
that has the interpretation of a fermionic number, allow us to
construct the following projectors on bosonic and fermionic
sectors:

\be
P_{\pm}~=~\frac{1}{2}\left(I\pm\Gamma_{N+1}\right)\label{proj}~,
\ee which are the generators of the usual projectors algebra

\be P_{a}P_{b}~=~\delta_{ab}P_{a}~. \ee
Commutation properties of
$P_{\pm}$ with $\Gamma$-matrices are easily seen to be
\bea
&P_{\pm}\Gamma_{I}~=~\Gamma_{I} P_{\mp}~,\nonumber \\
&P_{\pm}\Gamma_{N+1}~=~\pm \Gamma_{N+1}P_{\pm}~. \eea The way to go
back to ${\cal GR}({\rm d}, N)$ from a real Clifford algebra is
through \bea
&R_{I}~=~P_{+}\Gamma_{I} P_{-}~,\nonumber \\
~~&L_{I}~=~P_{-}\Gamma_{I}P_{+}~,\label{LR} \eea that yield
immediately the condition (\ref{cond1}) \bea
&R_{(I}L_{J)}~=~P_{+}\Gamma_{(I}P_{-}\Gamma_{J)}P_{+}~=~-2\delta_{IJ}P_{+}~\equiv~-2\delta_{IJ}{\bf I}_+ ~~,\\
&L_{(I}R_{J)}~=~P_{-}\Gamma_{(I}P_{+}\Gamma_{J)}P_{-}~=~-2\delta_{IJ}P_{-}~\equiv~-2\delta_{IJ}{\bf
I}_- ~~. \eea In this way, we have just demonstrated that
representations of ${\cal GR}(d, \, N)$ are in one-to-one
correspondence with real valued representations of Clifford
algebras, which will be classified in the following using
considerations of \cite {okubo}. To this end, let $M$ be an
arbitrary $d\times d$ real matrix and let us consider

\be
S~=~\sum_{A} \Gamma_{A}^{-1}M\Gamma_{A}~,
\ee
then

\be \forall~\Gamma_{B}\in
C(p,q)~,~~\Gamma_{B}^{-1}S\Gamma_{B}~=\sum_{A}\left(\Gamma_{B}\Gamma_{A}\right)^{-1}M\Gamma_{A}\Gamma_{B}~=\sum_{C}\Gamma_{C}^{-1}M\Gamma_{C}~=~S~,
\label{Mmatrix} \ee where we have used the property of
$\Gamma$-matrices

\be \Gamma_{A}\Gamma_{B}=\epsilon_{AB}\Gamma_{C}+\delta_{AB}I~.
\ee Equation (\ref{Mmatrix}) tells us that for all $\Gamma_{A}\in
C(p,q)$ there exists at least one $S$ such that
$\left[\Gamma_{A},S\right]=0$. Thus, by Shur's lemma, $S$ has to
be invertible (if not vanishing). It follows that any set of such
$M$ matrices defines a real division algebra. As a consequence of
a Frobenius theorem, three possibilities exist that we are going
to analyze.

\begin{enumerate}

\item{$\bf {Normal~representations~(N).}$}

The division algebra is generated by the identity only

\be S~=~\lambda I~,~~~\lambda\in R~. \ee

\item{$\bf {Almost~complex~representations~(AC).}$}

There exists a further division algebra real matrix $J$ such that
$J^2=-I$ and we have

\be S~=~\mu I+\nu J~,~~~\mu,\nu\in R~. \ee

\item{$\bf {Quaternionic~representations~(Q).}$}

Three elements $E_1,~E_2$ and $E_3$ satisfying quaternionic
relations

\be
E_iE_j~=~-\delta_{ij}E+\sum_{k=1}^{3}\epsilon_{ijk}E_k~,~~~i,j~=~1,2,3~,
\ee are present in this case. Thus it follows

\be S=\mu I+\nu E_1 +\rho E_2 +\sigma E_3~,~~\mu,\nu,\rho,\sigma~\in
R~. \ee
\end{enumerate}


The results about irreducible representations obtained in \cite
{okubo} for $C(p,q)$ are summarized in the table 1.

\begin{table}

\begin{center}

\begin{tabular}{|c|c|c|c|c|c|c|c|c| }

\hline

$\begin{array}{c} \\ p-q= \\ \\
\end{array}$  &
$~~~0~~~$&$~~~1~~~$&$~~~2~~~$&$~~~3~~~$&$~~~4~~~$&$~~~5~~~$&$~~~6~~~$&$~~~7~~~$ \\ \hline  \hline
$\begin{array}{c} \\ ~~Type~~ \\ \\
\end{array}$   & N  & N &
N &AC &Q &Q &Q &AC \\ \hline
$\begin{array}{c} \\ ~~Rep.~dim.~~ \\ \\
\end{array}$   & $~2^n~$  & $~2^n~$ & $~2^n~$ & $~~2^{n+1}~$ & $~~2^{n+1}~$ & $~~2^{n+1}~$ & $~~2^{n+1}~$ &  $~~2^{n+1}~$ \\ \hline
\end{tabular}\\[.1in]

\caption{Representation dimensions for $C(p,q)$ algebras. It appears $n~=~\left[ \frac{p+q}{2}\right]$ with [.] denoting here and in the following, the integer part. 
}
\end{center}

\label{repdim1}

\end{table}

The dimensions of irreducible representations are referred to
faithful ones except the $p-q=1,5$ cases where exist two
inequivalent representations of the same dimension, related to
each other by $\bar{\Gamma}_A=-\Gamma_A$. In order to obtain
faithful representations, the dimensions of those cases should be
doubled defining

\be
\widetilde{\Gamma}_{A}~=~\left(
\begin{array}{cc}
\Gamma_{A}& 0\\ 0& -\Gamma_{A}
\end{array}
\right)~.
\ee
Once the faithfulness has been recovered, we can say that a periodicity theorem holds, asserting that
\bea
C\left(p+8,~0\right)&~=~&C\left(p,~0\right)\otimes M_{16}\left(R\right)~, \\
C\left(0,~q+8\right)&~=~&C\left(0,~q\right)\otimes
M_{16}\left(R\right)~, \label{theorem1} \eea where
$M_{r}\left(R\right)$ stands for the set of all $r\times r$ real
matrices. Furthermore we have

\be C\left(p,p\right)~=~M_{r}\left(R\right)~,~~~r=2^n~.
\label{theorem2} \ee The structure theorems (\ref{theorem1}) and
(\ref{theorem2}) justify the restriction in table \ref{repdim1} to
values of $p-q$ from $0$ to $7$. As mentioned in \cite {Pash}, the
dimensions reported in table \ref{repdim1} can be expressed as
functions of the signature ($p,q$) introducing integer numbers
$k,l,m$ and $n$ such that \bea
q~=~8k+m~,~~0\leq m\leq 7~, \nonumber \\
p~=~8l+m+n,~~1\leq n\leq 8~,\label{pq}
\eea
where $n$ fix $p-q$ up to $l-k$ multiples of eight as can be seen from

\be
p-q~=~8(l-k)+n~,
\ee
while $m$ encode the $p,q$ choice freedom keeping $p-q$ fixed. Obviously $k$ and $l$ take into account the periodicity properties. The expression of irreducible representation dimensionalities reads

\be
d~=~2^{4k+4l+m}F(n)
\ee
where $F(n)$ is the Radon-Hurwitz function defined by

\be F(n)~=~2^{r}~,~~[log_{2}n]+1\geq r\geq [log_{2}n]~,~~r\in N~.
\ee

\vspace{0.5cm}

Turning back to ${\cal {GR}}(d,N)$ algebras, from (\ref{eta})
we deduce that we have to deal only with $C(N,1)$ case which means that
irreducible representation dimensions depend only on $N$ in the
following simple manner:

\be d=2^{4a}{\cal F}(b) \ee where $N=8a+b$ with $a$ and $b$
integer running respectively from $1$ to $8$ and from $0$ to
infinity. This result can be straightforwardly obtained setting
$p=1$ and $q=N$ in equations (\ref{pq}). Representation dimensions
obtained adapting the results of table 1 to the $C(N,1)$ case are
summarized in table 2.

\begin{table}

\begin{center}

\begin{tabular}{|c|c|c|c|c|c|c|c|c| }

\hline $\begin{array}{c} \\ b= \\ \\
\end{array}$  &
$~~~1~~~$&$~~~2~~~$&$~~~3~~~$&$~~~4~~~$&$~~~5~~~$&$~~~6~~~$&$~~~7~~~$&$~~~8~~~$
\\ \hline  \hline $\begin{array}{c} \\ ~~Type~~ \\ \\
\end{array}$   & N  & AC & Q &Q &Q &AC &N &N \\
\hline $\begin{array}{c} \\ ~~Rep.~dim.~~ \\ \\
\end{array}$   & $~~2^{4a}~~$  & $~~2\cdot 2^{4a}~$ & $~~4\cdot
2^{4a}~$ & $~~4\cdot 2^{4a}~$ & $~~8\cdot 2^{4a}~$ & $~~8\cdot
2^{4a}~$ & $~~8\cdot 2^{4a}~$ &  $~~8\cdot 2^{4a}~$ \\ \hline
\end{tabular}\\[.1in]

\caption{Representation dimensions for ${\cal {GR}}(d,N)$ algebras. 
}

\end{center}

\label{repdim2}

\end{table}


In what follows we focus our attention to the explicit
representations construction. First of all we enlarge the set of
linear mappings acting between ${\cal V}_L$ and ${\cal V}_R$,
namely ${{\cal M}_L}\oplus{{\cal M}_R}$ ( i.e. ${\cal {GR}}(d,N)$
), to ${{\cal U}_L}\oplus{{\cal U}_R}$ defining the enveloping
general real algebra

\be {\cal {EGR}}(d,N)={{\cal M}_L}\oplus{{\cal M}_R}\oplus{{\cal
U}_L}\oplus{{\cal U}_R}~. \ee As noticed before, we have the
possibility to construct elements of ${{\cal U}_L}$ and ${\cal
U}_R$ as products of alternating elements of ${{\cal M}_L}$ and
${\cal M}_R$ so that \bea
L_I R_J~,L_I R_J L_K R_L~,...~~\in{{\cal U}_R}~,\nonumber \\
R_I L_J~,R_I L_J R_K L_L~,...~~\in{{\cal U}_L}~, \eea but $L_I$
and $R_J$ comes from $C(N,1)$ through (\ref{LR}) so all the
ingredients are present to develop explicit representation of
${\cal {EGR}}(d,N)$ starting from Clifford algebra.

\vspace{0.3cm}

We focus now on the building of enveloping algebras
representations starting from Clifford algebras. Indeed we need to
divide into three cases.

\begin{enumerate}
\item {\bf Normal representations.} In this case basic definition of Clifford algebra
(\ref{cliffalg2}) suggests a way to construct a basis $\{ \Gamma \}$ by wedging
$\Gamma$ matrices

\be \{ \Gamma
\}~=~\{I,\Gamma^{I},\Gamma^{IJ},\Gamma^{IJK},..,\Gamma^{N+1}\}~,~~I<J<K...,
\ee where $\Gamma^{I,..,J}$ are to be intended as the
antisymmetrization of $\Gamma^{I}\cdot\cdot\Gamma^{J}$ matrices
otherwise denoted by $\Gamma^{[I}\cdots\Gamma^{J]}$ or
$\Gamma^{[N]}$ if the product involve $N$ elements. Dividing into
odd and even products of $\Gamma$ we obtain the sets \bea \{
\Gamma_e \} &~=~& \{~ {I}, \, \Gamma^{N+1}, \, \Gamma^{IJ}, \,
\Gamma^{IJ}\Gamma^{N+1}, \, \Gamma^{IJKL}, \dots ~\} ~,  \nonumber \\
\{ \Gamma_o \} &~=~& \{~  \Gamma^{I} , \, \Gamma^{I}\Gamma^{N+1},
\, \Gamma^{IJK} , \, \Gamma^{IJK}\Gamma^{N+1} \dots  ~\}~,
\label{evenodd} \eea respectively related to $\{{\cal M}\}$ and
$\{{\cal U}\}$ spaces. Projectors (\ref{proj}) have the key role
to separate left sector from right sector. In fact, for instance,
we have \bea
P_{+}\Gamma_{IJ}P_{+}&~=~&P_{+}\Gamma_{[I}\Gamma_{J]}P_{+}P_{+}~=~P_{+}\Gamma_{[I}P_{-}\Gamma_{J]}P_{+}~=~\nonumber \\
&~=~&P_{+}\Gamma_{[I}P_{-}P_{-}\Gamma_{J]}P_{+}~=~R_{[I}L_{J]}~\in\{{\cal
U}_{L}\}~, \eea and in a similar way
$P_{-}\Gamma^{IJ}P_{-}=L^{[I}R^{J]}~\in\{{\cal U}_{R}\},~
P_{+}\Gamma_{IJK}P_{-}=R_{[I}L_{J}R_{K]}~\in\{{\cal M}_{R}\}$, and
so on. Those remarks provide the following solution \bea \{{\cal
U}_R\}&~=~\{ \, P_{-}, \, P_{-}\Gamma^{IJ}P_{-} ~,..,~
P_{-}\Gamma^{[N]}P_{-}\}
\equiv&\{\,I_{{\cal V}_R},~L^{[I}R^{J]},\dots\}~, \nonumber \\
\{{\cal M}_R\} &~=~\{P_{+}\Gamma^{I}P_{-},..,~P_{+}\Gamma^{[N-1]}P_{-}\}~~~~~~\equiv&\{\,R^{I},~R^{[I}L^{J}R^{K]},\dots\} ~, \nonumber\\
\{{\cal
U}_L\}&~=~\{\,P_{+},~P_{+}\Gamma_{IJ}P_{+},..,~P_{+}\Gamma_{[N]}P_{+}\}\equiv&\{\,I_{{\cal V}_L},~R_{[I}L_{J]},\dots\}~, \nonumber \\
\{{\cal M}_L\}&~=~\{ P_{-}\Gamma_{I}
P_{+}~,...,~P_{-}\Gamma_{[N-1]}P_{+}
\}~~~~~\equiv&\{\,L_{I},~L_{[I}R_{J}L_{K]},\dots\}~, \nonumber \\
\label{repr} \eea which we will denote as $\wedge{\cal {GR}}(d,N)$
to remember that it is constructed by wedging $L_{I}$ and $R_{J}$
generators. Clearly enough, from each $\Gamma_{[I,..,J]}$ matrix
we get two elements of ${\cal {EGR}}(d,N)$ algebra as a
consequence of the projection. Thus we can say that in the normal
representation case, $C(N,1)$ is in one-two correspondence with
the enveloping algebra which can be identified by $\wedge{\cal
{GR}}(d,N)$. By the wedging construction in (\ref{repr}) naturally
arise p-forms that is useful to denote

\bea
f_{I}&~=~L_{I}~,~~~~~~~~~~~~~~~\hat{f}^{I}~=~&R^{I}~,\nonumber\\
f_{IJ}&~=~R_{[I}L_{J]}~,~~~~~~~~\hat{f}^{IJ}~=~&L^{[I}R^{J]}~,\nonumber\\
f_{IJK}&~=~L_{[I}R_{J}L_{K]} ~,~~\hat{f}^{IJK}~=~&R^{[I}L^{J}R^{K]}~,\nonumber\\
&\vdots~~~~~~~~~~~~~~~~~~~~~~~~~~~~~\vdots& \label{p-forms} \eea
The superfield components for the $\{~{\cal U}_L\}\oplus\{~{\cal
M}_L\}$ type superspace introduced in (\ref{cliffsupfield}), can
be expanded in terms of this normal basis as follows \bea
\phi_{i}^{~j}&~=~&\phi~\delta_{i}^{~j}+\phi^{IJ}(f_{IJ})_{i}^{~j}+..~\in \{~{\cal U}_L\}~,\nonumber\\
\psi_{\hat{\alpha}}^{~i}&~=~&\psi^{I}(f_I)_{\hat{\alpha}}^{~i}+\psi^{IJK}(f_{IJK})_{\hat{\alpha}}^{~i}+..~\in
\{~{\cal M}_L\}\label{boscliff} \eea according to the fact that
$f_{[even]}\in\{~{\cal U}_L\}$ and $f_{[odd]}\in\{~{\cal M}_L\}$.
We will refer to this kind of supefields as bosonic Clifford
algebraic ones because of the bosonic nature of the level zero
field. Similar expansion can be done for the $\{{\cal
U}_R\}\oplus\{{\cal M}_R\}$ type superspace where
$\hat{f}_{[even]}\in\{~{\cal U}_R\}$ and
$\hat{f}_{[odd]}\in\{~{\cal M}_R\}$ \bea
\phi_{\hat{\alpha}}^{~\hat{\beta}}&~=~&\phi~\delta_{\hat{\alpha}}^{~\hat{\beta}}+\phi^{IJ}(\hat{f}_{IJ})_{\hat{\alpha}}^{~\hat{\beta}}+..~\in \{~{\cal U}_R\}~,\nonumber\\
\psi^{~\hat{\alpha}}_{i}&~=~&\psi^{I}(\hat{f}_I)^{~\hat{\alpha}}_{i}+\psi^{IJK}(\hat{f}_{IJK})^{~\hat{\alpha}}_{i}+..~\in
\{~{\cal M}_R\}~.\label{fermcliff} \eea In the (\ref{fermcliff})
case we deal with a fermionic Clifford algebraic superfield
because the component $\phi$ is a fermion. For completeness we
include the remaining cases, namely $\{{\cal U}_R\}\oplus\{{\cal
M}_L\}$ superspace \bea \phi_{\hat{\alpha}}^{~\hat{\beta}}&~=~&\phi~\delta_{\hat{\alpha}}^{~\hat{\beta}}+\phi^{IJ}(\hat{f}_{IJ})_{\hat{\alpha}}^{~\hat{\beta}}+..~\in \{~{\cal U}_R\}~,\nonumber\\
\psi_{\hat{\alpha}}^{~i}&~=~&\psi^{I}(f_I)_{\hat{\alpha}}^{~i}+\psi^{IJK}(f_{IJK})_{\hat{\alpha}}^{~i}+..~\in
\{~{\cal M}_L\}~, \eea and $\{{\cal U}_L\}\oplus\{{\cal M}_R\}$
superspace \bea
\phi_{i}^{~j}&~=~&\phi~\delta_{i}^{~j}+\phi^{IJ}(f_{IJ})_{i}^{~j}+..~\in \{~{\cal U}_L\}~,\nonumber\\
\psi^{~\hat{\alpha}}_{i}&~=~&\psi^{I}(\hat{f}_I)^{~\hat{\alpha}}_{i}+\psi^{IJK}(\hat{f}_{IJK})^{~\hat{\alpha}}_{i}+..~\in
\{~{\cal M}_R\}~. \eea

\item{\bf Almost complex representations.} As already pointed out, those kind of representations contain one more generator $J$ with respect to normal representations so that to span all the space, the normal part, which is generated by wedging, is doubled to form the basis for the Clifford algebra

\be
\{\Gamma\}~=~\{I,J,\Gamma^{I},\Gamma^{I}J,\Gamma^{IJ},\Gamma^{IJ}J,..,\Gamma^{N+1},\Gamma^{N+1}J\}~.
\label{acrep} \ee Starting from (\ref{acrep}) it is
straightforward to apply considerations from (\ref{evenodd}) to
(\ref{repr}) to end with a ${\cal {EGR}}(d,N)$ almost complex
representation in 1-2 correspondence with the previous. Concerning
almost complex Clifford algebra superfields, it is important to
stress that we obtain irreducible representations only restricting
to the normal part.

\item{\bf Quaternionic representations.} Three more generators $E^{\alpha}$ satisfing

\be
\left[E^{\alpha},E^{\beta}\right]~=~2\epsilon^{\alpha\beta\gamma}E^{\gamma}
\ee have to be added to the normal part to give the following
quaternionic Clifford algebra basis

\be
\{\Gamma\}~=~\{I,E^{\alpha},\Gamma^{I},\Gamma^{I}E^{\alpha},\Gamma^{IJ},\Gamma^{IJ}E^{\alpha},..,\Gamma^{N+1},\Gamma^{N+1}E^{\alpha}\}~,
\label{Qrep} \ee which is four times larger than the normal part.
Again, repeating the projective procedure presented above,
generators of Clifford algebra (\ref{Qrep}) are quadrupled to
produce the ${\cal {EGR}}(d,N)$ quaternionic representation. Even
in this case only the normal part gives irreducible
representations for the Clifford algebra superfields.

\end{enumerate}

Notice that from the group manifold point of view, the presence of
the generator $J$ for the almost complex case and generators
$E^{\alpha}$ for the quaternionic one, separate the manifold into
sectors which are not connected by left or right group elements
multiplication giving rise to intransitive spaces. Division
algebra has the role to link those different sectors.

\vspace{0.3cm}

Finally we explain how to produce an explicit matrix
representation using a recursive procedure mentioned in \cite
{jim2} that can be presented in the following manner for the case
$N=8a+b$ with $a\geq~1$: \bea
L_{1}&~=~&i\sigma^{2}\otimes I_{b}\otimes I_{8a}~=~R_{1}~,\nonumber\\
L_{I}&~=~&\sigma^{3}\otimes (L_{b})_{I}\otimes I_{8a}~=~R_{I}~,~~1\leq I\leq b-1~,\nonumber\\
L_{J}&~=~&\sigma^{1}\otimes I_{b}\otimes (L_{8a})_{J}~=~R_{J}~,~~1\leq I\leq 8a-1~,\nonumber\\
L_{N}&~=~&I_{2}\otimes I_{b}\otimes
I_{8a}~=~-R_{N}~,\label{algorithm} \eea where $I_n$ stands for the
n-dimensional identity matrix while $L_{b}$ and $L_{8a}$ are
referred respectively to the cases $N=b$ and $N=8a$. Expressions
for the cases where $N\leq 7$ which are the starting points to
apply the algorithm in (\ref{algorithm}), can be found in appendix
A of \cite {jim1}.

\section{Relationships between different models}

It turns out that apparently different supermultiplets can be
related to each other using several operations.

\begin{enumerate}

\item leaving $N$ and $d$ unchanged, one can increase or decrease
the number of physical bosonic degrees of freedom (while
necessarily and simultaneously to decrease or increase the number
of auxiliary bosonic degrees of freedom) within a supermultiplet
by shifting the level of the superfield $\theta$-variables
expansion by mean of an automorphism on the superalgebra
representation space, commonly called automorphic duality (AD).

\item it is possible to reduce the number of supersymmetries mantaining fixed representation dimension (reduction).

\item the space-time coordinates can be increased preserving the supersymmetries (oxidation).

\item by a space-time compactification, supersymmetries can be eventually increased.

\end{enumerate}

These powerful tools can be combined together to discover new
supermultiplets or to relate the known ones. The first two points
will be analyzed the following paragraphs while for the last two
procedures, we remind to \cite{BHJP}, \cite{K} and references
therein.

\subsection{Automorphic duality transformations}

Until now, we encountered the following two types of
representation: the first one defined on ${\cal V}_L$ and ${\cal
V}_R$ superspace complemented with the second one, Clifford
algebraic superfields. In the latter case we observed that in
order to obtain irreducible representations, is needed a
restriction to normal representations or to their normal parts. If
we consider irreducible cases of Clifford algebraic superfields
then there exists the surprisingly possibility to transmute
physical fields into auxiliary ones changing the supermultiplet
degrees of freedom dynamical nature. The best way to proceed for
an explanation of the subject is to begin with the $N=1$ example
which came out to be the simplest. In this case only two
supermultiplets are present
\begin{itemize}
\item the scalar supermultiplet $(X,\psi)$ respectively composed of one
bosonic and one fermionic field arranged in the superfield \be
X(\tau,\theta)~=~X(\tau)+i\theta\psi(\tau)~, \ee with
transformation properties \bea
\delta_{Q}X&~=~&i\epsilon\psi~,\nonumber\\
\delta_{Q}\psi&~=~&\epsilon\partial_{\tau}X~; \label{smtransf} \eea

\item the spinor supermultiplet $(\xi,A)$ respectively composed of one
bosonic and one fermionic field arranged in the superfield \be
Y(\tau,\theta)~=~\xi(\tau)+\theta A(\tau)~, \ee with
transformation properties \bea
\delta_{Q}A&~=~&i\epsilon\partial_{\tau}\xi~,\nonumber\\
\delta_{Q}\xi&~=~&\epsilon A~.\label{spmtransf} \eea
\end{itemize}
The invariant Lagrangian for the scalar supermultiplet
transformations (\ref{smtransf}) \be {\cal L}~=~{\dot X}^2
+ig\psi{\dot {\psi}}~, \label{lag} \ee gives to the fields $X$ and
$\psi$ a dynamical meaning and offers the possibility to perform
an automorphic duality map that at the superfield level reads \be
Y(\tau,\theta)~=~-i{\cal D}X(\tau,\theta)~, \label{ad} \ee where
${\cal D}=\partial_{\theta}+i\theta\partial_{\tau}$ is the
superspace covariant derivative. At the component level (\ref{ad})
corresponds to the map upon bosonic components \be
X(\tau)~=~\partial_{\tau}^{-1}A(\tau)~, \label{map} \ee and
identification of fermionic ones. The mapping (\ref{map}) is
intrinsically not local but it can be implemented in a local way
both in the transformations (\ref{smtransf}) and in the Lagrangian
 (\ref{lag}) producing respectively equations (\ref{spmtransf}) and
the Lagrangian \be {\cal L}~=~A^2 +ig\psi{\dot {\psi}}~.
\label{lag2} \ee As a result we get that automorphic duality
transformations map $N=1$ supermultiplets into each other in a
local way, changing the physical meaning of the bosonic field $X$
from dynamical to auxiliary $A$ (not propagating) as is showed by
the Lagrangian (\ref{lag2}) invariant for (\ref{spmtransf})
transformations. Note that the auxiliary meaning of $A$ is already
encoded into (\ref{spmtransf}) transformations that enlighten on
the nature of the fields and consequently of the supermultiplet.

\vspace{0.5cm}

Let us pass to the analysis of the $N=2$ case making a link with the
considerations about representation theory discussed above. At the
$N=2$ level, we deal with a AC representation so, in order to
implement AD transformations, we focus on the normal part, namely
$\wedge{\cal {GR}}(d,N)$, defining the Clifford algebraic bosonic
superfield \bea
\phi^{~i}_{j}~&=&~\phi\delta^{~i}_{j}+\phi^{IJ}(f_{IJ})^{~i}_{j}~,
\nonumber\\
\psi^{~i}_{\hat{\alpha}}~&=&~\psi^{I}(f_{I})^{~i}_{\hat{\alpha}}~,
\label{exp} \eea constructed with the forms (\ref{p-forms}). Notice
that if we work in a $N$-dimensional space then the highest rank for
the forms is $N$. This is the reason why, writing (\ref{exp}), we
stopped at $f^{IJ}$ level. Some comments about transformation
properties. By comparing each level of the expansion, it is
straightforward to prove that superfields (\ref{boscliff}) transform
according to (\ref{CAtransf}) if the component fields
transformations are recognized to be \bea
\delta_{Q}\phi^{I_{1}{\cdot\cdot\cdot}I_{p_{even}}}&~=~&
-i\epsilon^{[I_{1}}\psi^{I_{2}{\cdot\cdot\cdot}I_{p_{even}}]}
+i(p_{even}+1)\epsilon_{J}\psi^{I_{1}{\cdot\cdot\cdot}I_{p_{even}}J}~,\nonumber\\
\delta_{Q}\psi^{I_{1}{\cdot\cdot\cdot}I_{p_{odd}}}&~=~&
-\epsilon^{[I_{1}}{\dot{\phi}}^{I_{2}{\cdot\cdot\cdot}I_{p_{odd}}]}
+i(p_{odd}+1)\epsilon_{J}{\dot{\phi}}^{I_{1}{\cdot\cdot\cdot}I_{p_{odd}}J}~.
\label{comptransf} \eea Therefore equations (\ref{comptransf}) for
the $N=2$ case read \bea
\delta_{Q}\phi~&=&~i\epsilon_{I}\psi^{I}~,\nonumber\\
\delta_{Q}\psi^{I}~&=&~-\epsilon^{I}{\dot{\phi}}
+2\epsilon_{J}{\dot{\phi}}^{IJ}~,\nonumber\\
\delta_{Q} \phi^{IJ}~&=&~-i\epsilon^{[I}\psi^{J]}~.
\label{N=2transf} \eea Once again, transformations
(\ref{N=2transf}) admit local AD maps between bosonic fields. In
order to discuss this ins a way that brings this discussion in line
with that of \cite{RussianApp}, we adhere to a convention that
list three numbers $(PB, \,PF, \, AB)$ where $PB$ denotes the
number of `propagating' bosonic fields, $AB$ denotes the number of
`auxiliary' bosonic fields and $PF$ denotes the number of
fermionic fields.
\newline $~~$ \newline \indent We
briefly list the resulting supermultiplets arising from the
dualization procedure.
\begin{itemize}
\item The AD map involving $\phi$ field \be
\phi(\tau)~=~\partial_{\tau}^{-1}A(\tau)~, \label{AD1} \ee yield a
$(1,2,1)$ supermultiplet whose transformations properties are \bea
\delta_{Q} A~&=&~i\epsilon_{I}\partial_{\tau}{\psi^{I}}~,\nonumber\\
\delta_{Q}\psi^{I}~&=&~-\epsilon^{I}A+2\epsilon_{J}{\dot{\phi}}^{IJ}~,\nonumber\\
\delta_{Q}
\phi^{IJ}~&=&~-i\epsilon^{[I}\psi^{J]}~.\label{AD1transf} \eea

\item By redefining the $\phi^{IJ}$ field
\be
\phi^{IJ}~=~\partial_{\tau}^{-1}B^{IJ}~,
\label{AD2}
\ee
another $(1,2,1)$ supermultiplet is obtained. Accordingly, we have
\bea
\delta_{Q}\phi~&=&~i\epsilon_{I}\psi^{I}~,\nonumber\\
\delta_{Q}\psi^{I}~&=&~-\epsilon^{I}{\dot{\phi}}+2\epsilon_{J}A^{IJ}~,\nonumber\\
\delta_{Q} A^{IJ}~&=&~-i\epsilon^{[I}\partial_{\tau}\psi^{J]}~.
\label{AD2transf}\eea \item Finally, if both redefinitions
(\ref{AD1}) and (\ref{AD2}) are adopted, then we are left with
$(0,2,2)$ spinor supermultiplet whose components behave as \bea
\delta_{Q} A~&=&~i\epsilon_{I}\partial_{\tau}\psi^{I}~,\nonumber\\
\delta_{Q}\psi^{I}~&=&~-\epsilon^{I}A+2\epsilon_{J}A^{IJ}~,\nonumber\\
\delta_{Q} A^{IJ}~&=&~-i\epsilon^{[I}\partial_{\tau}\psi^{J]}~.
\eea
\end{itemize}


It is important to stress that we can make redefinitions of
bosonic fields via AD maps that involve higher time derivatives.
For instance, by applying $\partial_{\tau}^2$ to the first
equation in (\ref{N=2transf}) together with the new field
introduction \be \phi~=~\partial_{\tau}^{-2}C~, \ee
transformations turn out to be free from nonlocal terms if AD for
the remaining fields \bea
\psi^{I}~&=&~i\partial_{\tau}^{-1}\xi^{I}\nonumber\\
\phi^{IJ}~&=&~\partial_{\tau}^{-1}D^{IJ} \eea are enforced. Thus
we end with \bea
\delta_{Q} C~&=&~-\epsilon_{I}\partial_{\tau}\xi^{I}\nonumber\\
\delta_{Q}\xi^{I}~&=&~i\epsilon^{I}C-2i\epsilon_{J}\dot{D}^{IJ}\nonumber\\
\delta_{Q} D^{IJ}~&=&~\epsilon^{[I}\xi^{J]}~.
\label{generalmulttransf} \eea From equations
(\ref{generalmulttransf}) one may argue that $C$ is auxiliary
while $D$ is physical. The point is which is the meaning of the
fields we started from? An invariant action from
(\ref{generalmulttransf}) is \be {\cal L}~=~C^{2}+ig\xi\dot\xi
+\dot D_{IJ}\dot D^{IJ}~, \ee so that going backward, we can
deduce the initial action \be {\cal
L}~=~\ddot{\phi}^{2}+ig\dot\psi\ddot\psi
+\ddot\phi_{IJ}\ddot\phi^{IJ}~. \ee

\vspace{0.5cm}

The examples above should convince any reader that Clifford
superfields are a starting point to construct a wider class of
representation by means of AD maps. Following this idea, one can
identify each supermultiplet with a correspondent root label
$(a_{1},..,a_{k})_{\pm}$ where $a_i \in Z$ are defined according
to \bea (\tilde\phi,~\tilde\psi^{I},~\tilde{\phi}^{IJ},\dots
)_{+}~=~(\partial^{-a_{0}}_{\tau}\phi,~\partial_{\tau}^{a_{1}}\psi^{I},
\partial_{\tau}^{-a_{2}}\phi^{IJ},\dots)_{+}~,\nonumber\\
(\tilde\psi,~\tilde\phi^{I},~\tilde{\psi}^{IJ},\dots
)_{-}~=~(\partial^{a_{0}}_{\tau}\psi,~\partial_{\tau}^{-a_{1}}\phi^{I},
\partial_{\tau}^{a_{2}}\psi^{IJ},\dots)_{-}~,\label{rootlabel}
\eea  and $\pm$ distinguish between Clifford superfields of
bosonic and fermionic type. For instance, the last supermultiplet
(\ref{generalmulttransf}), corresponds to the case
$(a_{0},a_{1},a_{2})_{\pm}=(2,-1,1)_{+}$. We name base superfield
the one with all zero in the root label $(0,\dots,0)_\pm$,
underling that in the plus (minus) case, this supermultiplet has
to be intended as the one with all bosons (fermion) differentiated
in the r.h.s. of variations. They are of particular interest the
supermultiplets whose roots label involve only $0$ and $1$. All
these supermultiplets form what we call root tree.

\subsection{Reduction}
It is shown in table 2 that  $N=8,7,6,5$ irreducible
representations have the same dimension. The same happens for the
$N=4,3$ cases. This fact reflect the possibility to relate those
supermultiplets via a reduction procedure. In order to explain how
this method works, consider a form $f_{I_{1}..I_{K}}$ and notice
that the indices $I_{1},..,I_{K}$ run on the number of
supersymmetries: reducing this number corresponds to diminishing
the components contained in the rank $k$ form. The remaining
components has to be rearranged into another form. For instance if
we consider a 3-rank form for the $N=8$ case then the number of
components is given by\footnote{
For the construction of $N=8$ supersymmetric mechanics, see \cite{Bell8};
the nonlinear chiral multiplet has been used in this
connection \cite{n8nonlin}, as well as in related tasks \cite{1}.}
$\left(\begin{array}{c}8\\3\end{array}\right)=56$ but, reducing to
the $N=7$ case and leaving invariant the rank, we get
$\left(\begin{array}{c}7\\3\end{array}\right)=35$ components. The
remaining ones can be rearranged in a 5-rank form. This means that
the maximum rank of Clifford superfield expansion is raised until
the irreducible representation dimension is reached. However the
right way to look at this rank enhancing is through duality. An
enlightening example will be useful. By a proper counting of
irreducible representation dimension for the ${\cal {EGR}}(8,8)$,
we are left with $\{{\cal U}_{L}\}\oplus\{{\cal M}_{L}\}$ type
Clifford algebraic superfield \bea
\phi_{ij}~&=&~\phi \delta_{ij}+\phi^{IJ}\left(f_{IJ}\right)_{ij}+\phi^{IJKL}\left(f_{IJKL}\right)_{ij}\nonumber\\
\psi_{{\hat {\alpha}}i}~&=&~\psi^{I}\left(f_{I}\right)_{{\hat
{\alpha}}i}+\psi^{IJK}\left(f_{IJK}\right)_{{\hat {\alpha}}i} \eea
where the 4-form has definite duality or, more precisely, the sign
in the equation \be \epsilon^{IJKLMNPQ}f_{MNPQ}~=~\pm
f^{IJKL}~,\ee has been chosen, halving the number of independent
components. In order to reduce to the $N=7$ case, we need to
eliminate all ``8" indices and this can be done by exploiting the
duality. For instance, $f_{I8}$ can disappear if transformed into
\be \epsilon^{IJKLMNP8}f_{P8}~=~\pm f^{IJKLMN}~. \ee This trick
adds the 6-rank to the expansion manifesting the enhancing
phenomenon previously discussed. Once the method is understood, it
is straightforward to prove that for the $N=7$ case, the proper
superfiels expression is \bea
\phi_{ij}~&=&~\phi\delta_{ij}+\phi^{IJ}\left(f_{IJ}\right)_{ij}+\phi^{IJKL}\left(f_{IJKL}\right)_{ij}+\phi^{IJKLMN}\left(f_{IJKLMN}\right)_{ij}\nonumber\\
\psi_{i{\hat j}}~&=&~\psi^{I}\left(f_{I}\right)_{i{\hat
j}}+\psi^{IJK}\left(f_{IJK}\right)_{i{\hat
j}}+\psi^{IJKLM}\left(f_{IJKLM}\right)_{i{\hat
j}}+\nonumber\\&&+\psi^{IJKLMPQ}\left(f_{IJKMNPQ}\right)_{i{\hat
j}}~. \eea The explicit reduction procedure for $N\leq 8$ can be
found in \cite {jim0} and summarized in the following tables:

\begin{table}

\begin{center}

\begin{tabular}{|c|c|c| }

\hline ~${\cal {EGR}}(d,N)$~&$\begin{array}{c} \\ \Lambda{\cal
{GR}}(d,N)~basis \\ \\ \end{array}$&$Division~Structure$
\\ \hline  \hline ${\cal {EGR}}(4,4)$&$\begin{array}{c} \\ \{{\cal U}_{L}\}~=~\{I,f_{IJ},{\cal E}^{\hat{\mu}},f_{IJ}{\cal E}^{\hat{\mu}}\}\\ \\ \{{\cal M}_{L}\}~=~\{f_{I},f_{I}\hat{\cal E}^{\hat{\mu}}\} \\ \\ \end{array}$ & $~~~~~{\cal E}^{\hat{\mu}},~\hat{\cal E}^{\hat{\mu}}~~~~$ \\
\hline ${\cal {EGR}}(4,3)$&$\begin{array}{c} \\ \{{\cal U}_{L}\}~=~\{I,f_{IJ},{\cal E}^{\hat{\mu}},f_{IJ}{\cal E}^{\hat{\mu}}\}\\ \\ \{{\cal M}_{L}\}~=~\{f_{I},f_{I}\hat{\cal E}^{\hat{\mu}},f_{IJK},f_{IJK}\hat{\cal E}^{\hat{\mu}}\}\\ \\ \end{array}$ & $~~~~~{\cal E}^{\hat{\mu}},~\hat{\cal E}^{\hat{\mu}}~~~~$ \\
\hline
\end{tabular}\\[.1in]

\caption{${\cal {EGR}}(4,4)$ and its reduction: algebras
representation in terms of forms and division algebra. Here and in
the following table, the generators ${\cal
E}^{\hat{\mu}},~\hat{\cal E}^{\hat{\mu}}$ are respectively the +
and - projections of the quaternionic division algebra generators
in the Clifford space. The same projection on complex structure
originate $D,~\hat{D}$.}
\end{center}

\label{red2}

\end{table}

\begin{table}

\begin{center}

\begin{tabular}{|c|c|c| }

\hline ~${\cal {EGR}}(d,N)$~ & $\begin{array}{c} \\ \Lambda{\cal
{GR}}(d,N)~basis \\ \\ \end{array}$ & $Division~Structure$
\\ \hline  \hline ${\cal {EGR}}(8,8)$&$\begin{array}{c} \\ \{{\cal U}_{L}\}~=~\{I,f_{IJ},f_{IJKL}\}\\ \\ \{{\cal M}_{L}\}~=~\{f_{I},f_{IJK}\} \\ \\ \end{array}$ & ~~~~~I~~~~~\\
\hline ${\cal {EGR}}(8,7)$&$\begin{array}{c} \\ \{{\cal U}_{L}\}~=~\{I,f_{IJ},f_{IJKL},f_{IJKLMN}\}\\ \\ \{{\cal M}_{L}\}~=~\{f_{I},f_{IJK},f_{[5]},f_{[7]}\}\\ \\ \end{array}$ &~~~~~I~~~~~  \\
\hline ${\cal {EGR}}(8,6)$&$\begin{array}{c} \\ \{{\cal U}_{L}\}~=~\{I,f_{I7},f_{IJ},f_{IJK7},f_{IJKL},f_{[5]7},f_{[6]}\}\\ \\ \{{\cal M}_{L}\}~=~\{f_{7},f_{I},f_{IJ7},f_{IJK},,f_{[4]7},f_{[5]},f_{[6]7}\} \\ \\ \end{array}$ &$D~=~f_{7}$  \\
\hline ${\cal {EGR}}(8,5)$&$\begin{array}{c} \\ \{{\cal
U}_{L}\}~=~\{I,f_{67},f_{I6},f_{I7},f_{IJ},f_{IJ67},f_{IJK7},\\
\\~~~~~~~~~~~~~~f_{IJKL},f_{IJKL67},f_{[5]7},f_{[5]6}\}\\ \\
\{{\cal
M}_{L}\}~=~\{f_{7},f_{6},f_{I},f_{I67},f_{IJ7},f_{IJ6},f_{IJK},\\
\\~~~~~~~~~~~~~~f_{IJk67},f_{[4]7},f_{[4]6},f_{[5]},f_{[5]67},\} \\ \\ \end{array}$
& $\begin{array}{c} \\ {\cal
E}^{\hat{\mu}}~=~(f_{67},f_{[5]6},f_{[5]7})\\ \\ \hat{\cal
E}^{\hat{\mu}}~=~(f_{7},f_{6},f_{[5]67}) \\ \\
\end{array}$\\
\hline
\end{tabular}\\[.1in]

\caption{${\cal {EGR}}(8,8)$ and its reductions. Here the
subscript [n] is used in place of n anticommuting indices. }

\end{center}

\label{red1}

\end{table}

\section{Applications}

\subsection{Spinning particle}
Before we begin a detailed analysis of spinning particle system it
is important to understand what a spinning particle is. Early
models of relativistic particle with spin involving only commuting
variables can be divided into the two following classes:
\begin{itemize}
\item vectorial models, based upon the idea of extending Minkowski
space-time by vectorial internal degrees of freedom;
\item spinorial models, characterized by the enhancing of
configuration space using spinorial commuting variables.
\end{itemize}
These models lack of the following important requirement: after
first quantization, they never produce relativistic Dirac
equations. Moreover, in the spinorial cases, a tower of all
possible spin values appear in the spectrum. Further progress in
the development of spinning particle descriptions was achieved by
the introduction of anticommuting variables to describe internal
degrees of freedom \cite{M}. This idea stems from the classical
limit ($h\rightarrow 0$) formulation of Fermi systems \cite{C},
the so called 'pseudoclassical mechanics' referring to the fact
that it is not a ordinary mechanical theory because of the
presence of Grassmannian variables. By means of pseudoclassical
approach, vectorial and spinorial models can be generalized to
'spinning particle' and 'superparticle' models, respectively. In
the first case, the extension to superspace ($x_{\mu},
\theta_{\mu}, \theta_{5}$) is made possible by a pseudovector
$\theta_{\mu}$ and a pseudoscalar $\theta_{5}$ \cite{BCL},
\cite{BM}. The presence of vector index associated with
$\theta$-variables implies the vectorial character of the model.

In the second case, spinorial coordinates are considered, giving
rise to ordinary superspace approach whose underlying symmetry is
the super Poincar\' e group (eventually extended) \cite{C}. The
superparticle is nothing but a generalization of relativistic
point particle to superspace.

It turns out that after first quantization, the spinning particle
model produced Dirac equations and all Grassmann variables are
mapped into Clifford algebra generators. Superfields that take
values on this kind of quantized superspace are precisely Clifford
algebraic superfields described in the previous paragraphs. On the
other side, a superspace version of Dirac equation arises from
superparticle quantization. Moreover, $\theta$-variables are still
present in the quantized version.

To have a more precise idea, we spend a few words discussing the
Barducci-Casalbuoni-Lusanna  model \cite{BCL} which is one of the
first works on pseudoclassical model. As already mentioned, it is
assumed the configuration space to be described by ($x_{\mu},
\theta_{\mu}, \theta_{5}$). The Lagrangian of the system \be {\cal
L}_{BCL}~=~-m\sqrt{\left({\dot
x}^{\mu}-\frac{i}{m}\theta^{\mu}\dot{\theta_{5}}\right)\left({\dot
x}_{\mu}-\frac{i}{m}\theta_{\mu}\dot{\theta_{5}}\right)}-\frac{i}{2}\theta_{\mu}\dot{\theta}^{\mu}-\frac{i}{2}\theta_{5}\dot{\theta}_{5}
\ee is invariant under the transformations \bea
\delta x_{\mu}~&=&~-\epsilon_{\mu}a\theta_{5}+\epsilon_{5}b\theta_{\mu}~,\nonumber\\
\delta\theta_{\mu}~&=&~\epsilon_{\mu}~,\nonumber\\
\delta\theta_{5}~&=&~\epsilon_{5}~. \eea and produces the
equations of
motion \bea p^2-m^2~=~0~,\nonumber\\
p_{\mu}\theta^{\mu}-m\theta_{5}~=~0~,\label{eqmot} \eea after a
canonical analysis. These equations (\ref{eqmot}) are classical
limits of Klein-Gordon and Dirac equations, respectively.
Moreover, the first quantization maps $\theta$-variables into
Clifford algebra generators \bea
\theta_{\mu}&\rightarrow&\gamma_{\mu}\gamma_{5}~~(pseudovector)~,\nonumber\\
\theta_{5}&\rightarrow&\gamma_{5}~~(pseudoscalar)~, \eea so that
equations (\ref{eqmot}) exactly reproduce relativistic quantum
behavior of a particle with spin. Even if it is not manifest, it
is possible to find a particular direction in the
$(\theta_{\mu},\theta_{5})$ space along which the theory is
invariant under the following localized supersymmetry
transformation \bea
\delta x_{\mu}~&=&~2\frac{i}{m^{2}}\epsilon_{5}(\tau)P_{\mu}\theta_{5}(\tau)-\frac{i}{m}\epsilon_{5}(\tau)\theta_{\mu}~,\nonumber\\
\delta\theta_{\mu}~&=&~\frac{1}{m}\epsilon_{5}(\tau)P_{\mu}~,\nonumber\\
\delta\theta_{5}~&=&~\epsilon_{5}(\tau)~, \eea opening the way to
supergravity. Basic concepts on the extension to minimal
supergravity-coupled model can be found in \cite{BDVH} . Here the
proposed action is a direct generalization of 1-dimensional
general covariant free particle to include the spin; for the
first-order formalism in the massless case we have \be S~=~\int
d\tau\{P^{\mu}\dot{X}_{\mu}-\frac{1}{2}eP^{2}-\frac{i}{2}\psi^{\mu}\dot{\psi}_{\mu}-\frac{i}{2}\chi\psi^{\mu}P_{\mu}\}~,\label{nomassact}
\ee with the local invariances \bea &\delta\psi^{\mu}~=~\epsilon
(\tau)P^{\mu}~,~~~\delta\phi^{\mu}~=~i\epsilon(\tau)\psi^{\mu}~,~~~\delta
P^{\mu}~=~0~,\nonumber\\
&\delta
e~=~i\epsilon(\tau)\chi~,~~~\delta\chi~=~2\dot{\epsilon}(\tau)~,\eea
corresponding to pure supergravity transformations as is shown
calculating the commutators \bea
&\left[\delta_{\epsilon_{1}},\delta_{\epsilon_{2}}\right]X^{\mu}~=~\xi\dot{X}^{\mu}+i\tilde{\epsilon}\psi^{\mu}~,\nonumber\\
&\left[\delta_{\epsilon_{1}},\delta_{\epsilon_{2}}\right]\psi^{\mu}~=~\xi\dot{\psi}^{\mu}+\tilde{\epsilon}
P^{\mu}~,\nonumber\\
&\left[\delta_{\epsilon_{1}},\delta_{\epsilon_{2}}\right]e~=~\xi\dot{e}+\dot{\xi}e+i\tilde{\epsilon}\chi~,\nonumber\\
&\left[\delta_{\epsilon_{1}},\delta_{\epsilon_{2}}\right]\chi~=~\xi\dot{\chi}+\dot{\xi}\chi+2\dot{\tilde{\epsilon}}~,
\label{sugracomm} \eea where \bea
\xi~=~2ie^{-1}\epsilon_{2}\epsilon_{1}~,\nonumber\\
\tilde{\epsilon}~=~-\frac{1}{2}\xi\chi~. \eea In fact, the r.h.s.
of (\ref{sugracomm}) describes both general coordinate and local
supersymmetry transformations.

In order to produce a mass-shell condition, the massive version of
the above model require the presence of a cosmological term in the
action \be S~=~-\frac{1}{2}\int d\tau em^{2} \label{cosmterm}\ee
that, in turn, imply the presence of an additional anticommuting
field $\psi_{5}$, transforming through \be \delta
\psi_{5}~=~m\tilde{\epsilon}~,\label{psi5}\ee to construct terms
that restore the symmetries broken by (\ref{cosmterm}). The
complete action describing the massive spinning particle version
minimally coupled to supergravity multiplet turns out to be \be
S~=~\int d\tau
\left[P^{\mu}\dot{X}_{\mu}-\frac{1}{2}e(P^{2}+m^{2})
-\frac{i}{2}(\psi^{\mu}\dot{\psi}_{\mu}+\psi_{5}\dot{\psi}_{5})
-\frac{i}{2}\chi(\psi^{\mu}
P_{\mu}+m\psi_{5})\right]~.\label{massact}\ee The second-order
formalism for the massless and massive model follow
straightforwardly from actions (\ref{massact}) and
(\ref{nomassact}) eliminating the $P$ fields using their equations
of motion.

An advance on this line of research yielded the on-shell
N-extension \cite{HPPT}. However a satisfactory off-shell
description with arbitrary N require the ${\cal GR}(d, \, N)$
approach. In the paragraphs below we describe in detail how this
construction is worked out.

\vspace{0.5cm}

\subsubsection{Second-order
formalism for spinning particle with rigid N-extended
supersymmetry }

The basic objects of this model are Clifford algebraic bosonic
superfields valued in $\{{\cal U}_L\}\oplus\{{\cal M}_L\}$
superspace with transformations (\ref{CAtransf}). One can easily
check that the action \be S~=~\int d\tau
\{(\partial_{\tau}(\phi_{1})_{i}^{~j})(\partial_{\tau}(\phi_{1})_{j}^{~i})
+i(\psi_{1})_{i}^{~\hat{\alpha}}\partial_{\tau}(\psi_{1})_{\hat{\alpha}}^{~i}
\} \label{act}\ee is left unchanged by (\ref{CAtransf}). The next
step consists in separating the physical degrees of freedom in
$\left((\phi_{1})_{i}^{~j},~(\psi_{1})_{\hat{\alpha}}^{~i}\right)$
from nonphysical ones. For the bosonic superfield, valued in
$\{{\cal U}_L\}$, we separate the trace from the remaining
components \bea
(\phi_{1})_{i}^{~j}~=~X \delta_{i}^{~j}+\tilde{\phi}_{i}^{~j}~,\nonumber\\
X~=~\phi^{~i}_{i}~,~~\tilde{\phi}^{~i}_{i}~=~0~, \eea and perform
an AD transformation on tilded components \be
\tilde{\phi}_{i}^{~j}~=~\partial_{\tau}^{-1}{\cal F}_{i}^{~j}~,\ee
to end with the decomposition \be \phi_{i}^{~j}~=~X
\delta_{i}^{~j}+\partial_{\tau}^{-1}{\cal F}_{i}^{~j}~,\ee
constrained by the equation \be {\cal
F}_{i}^{~i}~=~0~.\label{dec1}\ee The field component $X$ can be
interpreted as the spinning particle bosonic coordinate in a
background space. Nothing forbids us from considering $D$
supermultiplet of this kind that amount to add a $D$-dimensional
background index $\mu$ to the superfields \bea
(\phi_{1})_{i}^{~j}\rightarrow(\phi_{1}^{\mu})_{i}^{~j}~,\nonumber\\
(\psi_{1})_{\hat{\alpha}}^{~i}\rightarrow(\psi_{1}^{\mu})_{\hat{\alpha}}^{~i}~.\eea
In this way the dimension of the background space has no link
neither with the number of supersymmetries nor with representation
dimension. However, to simplify the notation, background index
will be omitted. Transformations involving the fields defined in
(\ref{dec1}) reads \bea \delta_{Q}
X~&=&~-\frac{1}{d}i\epsilon^{I}(R_{I})_{i}^{~\hat
{\alpha}}(\psi_{1})_{\hat{\alpha
}}^{~i}~,\nonumber\\
\delta_{Q}{\cal F}_{i}^{~j}~&=&~'
i\epsilon^{I}(R_{I})_{i}^{~\hat {\alpha}}\partial_{\tau}(\psi_{1})_{\hat {\alpha}}^{~j}~,\nonumber\\
\delta_{Q}(\psi_{1})_{\hat
{\alpha}}^{~i}~&=&~\epsilon^{I}(L_{I})_{\hat {\alpha}}^{~j}{\cal
F}_{j}^{~i}+\epsilon^{I}(L_{I})_{\hat
{\alpha}}^{~i}\partial_{\tau}X~.\label{transf1}\eea Even in the
fermionic case, we need that only the lowest component in the
expansion (\ref{boscliff}) has physical meaning so that the higher
level components can be read as auxiliary ones by means of AD map.
Fermionic superfield components happen to be distributed in the
following manner \be (\psi_{1})_{\hat
{\alpha}}^{~i}~=~\psi^{I}(L_{I})_{\hat
{\alpha}}^{~i}+{\tilde{\psi}}_{\hat
{\alpha}}^{~i}~=~\psi^{I}(L_{I})_{\hat {\alpha}}^{~i}+\mu_{\hat
{\alpha}}^{~i}~,\label{dec2}\ee where
$\psi^{I}=\frac{1}{d}(R_{I})_{i}^{~\hat {\alpha}}(\psi_{1})_{\hat
{\alpha}}^{~i}$ and the fermionic superfield $\mu_{\hat
{\alpha}}^{~i}$ obey the constraint equation \be
(R_{I})_{i}^{~\hat {\alpha}}\mu_{\hat
{\alpha}}^{~i}~=~0~.\label{constr}\ee After the substitution of
the new component fields (\ref{dec2}), transformations
(\ref{transf1}) became \bea
&&\delta_{Q}X~=~-\frac{1}{d}i\epsilon^{I}(R_{I})_{i}^{~\hat
{\alpha}}(L_{J})^{~i}_{\hat {\alpha}}\psi^{J}~,\label{eq1}\\
&&\delta_{Q}{\cal F}_{i}^{~j}~=~i\epsilon^{I}(R_{I})_{i}^{~\hat
{\alpha}}(L_{J})^{~j}_{\hat {\alpha}}\partial_{\tau}\psi^{J}
-i\epsilon^{I}(R_{I})_{i}^{~\hat {\alpha}}\partial_{\tau}\mu_{\hat {\alpha}}^{~j}~,\label{eq2}\\
&&(L_{I})_{\hat
{\alpha}}^{~i}\delta_{Q}\psi^{I}+\delta_{Q}\mu_{\hat
{\alpha}}^{~i}~=~\epsilon^{I}(L_{I})_{\hat{\alpha}}^{~j}{\cal
F}_{j}^{~i}+\epsilon^{I}(L_{I})_{\hat
{\alpha}}^{~i}\partial_{\tau}X~.\label{eq3} \eea where we used
(\ref{constr}) to obtain (\ref{eq1}). Equations (\ref{eq1}) and
(\ref{eq2}) can be simplified into \bea
&&\delta_{Q}X~=~i\epsilon^{I}\psi_{I}~,\nonumber\\
&&\delta_{Q}{\cal
F}_{i}^{~j}~=~i\epsilon^{I}(f_{IJ})_{i}^{~j}\partial_{\tau}\psi^{I}-i\epsilon^{I}(R_{I})_{i}^{~\hat
{\alpha}}\partial_{\tau}\mu_{\hat
{\alpha}}^{~j}~,\label{USPMtransf} \eea if one notice that \bea
(R_{I})_{i}^{~\hat {\alpha}}(L_{J})^{~i}_{\hat
{\alpha}}~=~-d\delta_{IJ}~,\nonumber\\(R_{I})_{i}^{~\hat
{\alpha}}(L_{J})^{~j}_{\hat {\alpha}}~=~(f_{IJ})^{~j}_{i} \eea
while the equation (\ref{eq3}) need more care. In order to
separate the variation of $\psi^{I}$ and $\mu_{\hat
{\alpha}}^{~i}$, we multiply by $(R_{J})_{i}^{~\hat {\alpha}}$ to
eliminate the $\mu_{\hat {\alpha}}^{~i}$ contribution thanks to
(\ref{constr}). As a result we get \bea
-d\delta_{Q}\psi_{J}~&=&~\epsilon^{I}(f_{JI})_{i}^{~j}{\cal
F}_{j}^{~i}-d\epsilon_{J}\partial_{\tau}X\nonumber\\
&\Downarrow &
\nonumber\\
\delta_{Q}\psi_{I}~&=&~\epsilon_{I}\partial_{\tau}X-\frac{1}{d}\epsilon^{J}(f_{IJ})_{i}^{~j}{\cal
F}_{j}^{~i}~.\label{transf3} \eea Substituting back
(\ref{transf3}) into (\ref{eq3}) we finally have \bea
\delta_{Q}\mu_{\hat {\alpha}}^{~i}~&=&-(L_{I})_{\hat
{\alpha}}^{~i}\left[\epsilon^{I}\partial_{\tau}X-\frac{1}{d}\epsilon^{J}(f^{I}_{~J})_{i}^{~j}{\cal
F}_{j}^{~i}\right]+\epsilon^{I}(L_{I})_{\hat {\alpha}}^{~j}{\cal
F}_{j}^{~i}+\epsilon^{I}(L_{I})_{\hat
{\alpha}}^{~i}\partial_{\tau}X\nonumber\\
~&=&~\left[\epsilon^{I}(\hat{f}_{I})_{\hat
{\alpha}}^{~j}+\frac{1}{d}\epsilon^{J}(\hat{f}_{I})_{\hat
{\alpha}}^{~i}(f^{I}_{~J})_{i}^{~j}\right]{\cal
F}_{j}^{~i}~.\label{transf4} \eea The supermultiplet
$\left(X,{\cal F}_{i}^{~j},\psi_{I},\mu_{\hat
{\alpha}}^{~i}\right)$ together with transformations
(\ref{USPMtransf}), (\ref{transf3}) and (\ref{transf4}), is called
'universal spinning particle multiplet' (USPM). Acting with the
maps (\ref{dec1}) and (\ref{dec2}) on the action (\ref{act}) we
obtain the USPM invariant action \be S~=~\int d \tau
\{d(\partial_{\tau} X \partial_{\tau} X
-i\psi_{I}\partial_{\tau}\psi_{I})+{\cal F}_{i}^{~j}{\cal
F}_{i}^{~j}+i\mu_{i}^{~\hat{\alpha}}\partial_{\tau}\mu^{~i}_{\hat{\alpha}}\}
\ee that represent the second-order approach to the spinning
particle problem with global supersymmetry. A remarkably
difference between the AD presented in section (2.1) and the AD
used to derive USPM resides in the fact that in the latter case we
map $\tilde{\phi}_{i}^{~j}$ and
${\tilde{\psi}}_{\hat{\alpha}}^{~i}$ which are Clifford algebraic
superfield while in the previous we work at the component level.
Finally, it is important to keep in mind that the superfields
$\tilde{\phi}_{i}^{~j}$ and ${\tilde{\psi}}_{\hat{\alpha}}^{~i}$
take values on the normal part of the enveloping algebra which is
equivalent to say that they can be expanded on the basis
(\ref{p-forms}).

\subsubsection{First-order formalism for spinning particle with rigid N-extended supersymmetry}

To formulate a first-order formalism, one more  fermionic
supermultiplet is required. This time the superfields
$\left((\phi_{2})_{i}^{~j},~(\psi_{2})^{~\hat{\alpha}}_{i}\right)$,
valued in $\{{\cal U}_L\}\oplus\{{\cal M}_R\}$ superspace,
transform according to \bea
&\delta_{Q}(\phi_{2})_{i}^{~j}~=~-i\epsilon^I{(L_I)}^{~j}_{\hat{\alpha}}\partial_{\tau}(\psi_{2})^{~\hat{\alpha}}_{i} \nonumber \\
&\delta_{Q}(\psi_{2})^{~\hat{\alpha}}_{i}~=~\epsilon^I{(R_I)}^{~\hat{\alpha}}_{j}(\phi_{2})_{i}^{~j}~.
\eea The expansions needed turns out to be \bea
&(\phi_{2})_{i}^{~j}~=~P
\delta_{i}^{~j}+{\cal G}_{i}^{~j}~,~~~{\cal G}_{i}^{~i}~=~0\nonumber\\
&(\psi_{2})^{~\hat{\alpha}}_{i}~=~\bar{\psi}^{I}(R_{I})^{~\hat
{\alpha}}_{i}+{{\cal X}}^{~\hat
{\alpha}}_{i}~,~~~(L_I)^{~i}_{\hat{\alpha}}{{\cal X}}^{~\hat
{\alpha}}_{i}~=~0 \label{dec3} \eea that bring us to the
trasformations \bea
\delta_{Q}P&~=~&i\epsilon^{I}\bar{\psi}_{I}~,\nonumber\\
\delta_{Q}{\cal G}_{i}^{~j}&~=~&-i\partial_{\tau}\epsilon_{J}(
\hat{f}_{IJ})_{i}^{~j}\bar{\psi}_{I}-i\epsilon^{K}(L_{K})_{\hat{\alpha}}^{~j}{\cal X}_{i}^{~\hat{\alpha}}~,\nonumber\\
\delta_{Q}\bar{\psi}_{I}&~=~&\epsilon_{I}P+d^{-1}\epsilon^{J}(
\hat{f}_{JI})_{j}^{~i}{\cal G}_{i}^{~j}~,\nonumber\\
\delta_{Q}{\cal X}_{i}^{~\hat{\alpha}}&~=~&-d^{-1}\epsilon^{J}(
\hat{f}^{I})_{i}^{~\hat{\alpha}}( \hat{f}_{JI})_{j}^{~i}{\cal
G}_{i}^{~j}+\epsilon^{I}( \hat{f}_{I})_{j}^{~\hat{\alpha}}{\cal
G}_{i}^{~j} .\eea Here, the scalar supermultiplet
$\left((\phi_{1})_{i}^{~j},~(\psi_{1})_{\hat{\alpha}}^{~i}\right)$
has to be treated in the following different way:
 the off-trace
superfield $\mu_{\hat{\alpha}}^{~i}$ undergoes an AD \be
\mu_{\hat{\alpha}}^{~i}\rightarrow
\partial_{\tau}^{-1}\Lambda_{\hat{\alpha}}^{~i}~, \ee
that slightly changes the variation (\ref{USPMtransf}),
(\ref{transf3}) and (\ref{transf4}) into \bea
&&\delta_{Q}X~=~i\epsilon^{I}\psi_{I}~,\nonumber\\
&&\delta_{Q}{\cal
F}_{i}^{~j}~=~i\epsilon^{I}(f_{IJ})_{i}^{~j}\partial_{\tau}\psi^{I}-i\epsilon^{I}(R_{I})_{i}^{~\hat
{\alpha}}\Lambda_{\hat
{\alpha}}^{~j}~,\nonumber\\&&\delta_{Q}\psi_{I}~=~\epsilon_{I}\partial_{\tau}X+\frac{1}{d}\epsilon^{J}(f_{IJ})_{i}^{~j}{\cal
F}_{j}^{~i}~, \nonumber\\&&\delta_{Q}\Lambda_{\hat
{\alpha}}^{~i}~=~\left[\epsilon^{I}(L_{I})_{\hat
{\alpha}}^{~j}-\frac{1}{d}\epsilon^{J}(\hat{f}_{I})_{\hat
{\alpha}}^{~i}(f^{I}_{~J})_{i}^{~j}\right]\partial_{\tau}{\cal
F}_{j}^{~i}~.\label{transf6} \eea The action can be thought as the
sum of two separated pieces \bea S_{P^{2}}&~=~&\frac{1}{d}\int d
\tau\{
\left(\phi_{2}\right)_{i}^{~j}\left(\phi_{2}\right)_{j}^{~i}+i\left(\psi_{2}\right)_{\hat{\alpha}}^{~i}\partial_{\tau}\left(\psi_{2}\right)_{i}^{~\hat{\alpha}}\}~,\nonumber\\
S_{PV}&~=~&\frac{1}{d}\int d \tau\{
\left(\phi_{2}\right)_{i}^{~j}\partial_{\tau}\left(\phi_{1}\right)_{j}^{~i}+i\left(\psi_{2}\right)_{i}^{~\hat{\alpha}}\partial_{\tau}\left(\psi_{1}\right)_{\hat{\alpha}}^{~i}\}~,\eea
so that, in analogy with the free particle description where the
Lagrangian has the form \be {\cal L}~=~PV-\frac{1}{2}P^{2}\ee we
consider \be S~=~S_{PV}-\frac{1}{2}S_{P^{2}}~,\label{SPact}\ee as
the correct first-order free spinning particle model. By
eliminating the fermionic supermultiplet superfields by their
equations of motion, we fall into the second-order description. It
is clear that the fermionic supermultiplet is nothing but the
conjugated of USPM. After (\ref{dec3}) substitution the proposed
action (\ref{SPact}) assume the final aspect \bea S_{sp}~=~\int d
\tau\{&& P\partial_{\tau}X+\frac{1}{d}{\cal G}_{i}^{~j}{\cal
F}_{j}^{~i}-i\bar{\psi}^{I}\partial_{\tau}\psi_{I}+\frac{i}{d}{\cal X}_{i}^{~\hat{\alpha}}\Lambda^{~i}_{\hat{\alpha}}\nonumber\\
&&-\frac{1}{2}P^{2}-\frac{1}{2d}{\cal G}_{i}^{~j}{\cal
G}_{j}^{~i}-\frac{i}{2}\bar{\psi}^{I}\partial_{\tau}\bar{\psi}_{I}-\frac{i}{2d}{\cal
X}_{i}^{~\hat{\alpha}}\partial_{\tau}{\cal
X}_{\hat{\alpha}}^{~i}~~\}~,\eea where the auxiliary superfields
$P^{2},~{\cal F}_{i}^{~j},~\Lambda^{~i}_{\hat{\alpha}},~{\cal
G}_{i}^{~j}$ and ${\cal X}_{i}^{~\hat{\alpha}}$ are manifest.

\subsubsection{Massive theory}

The massive theory is obtained by adding to the previous first and
second-order actions the appropriate terms where it figures an
additional supermultiplet
$\left(\hat{\psi}_{i}^{~\hat{\alpha}},~\hat{G}_{i}^{~j}\right)~,$
which is fermionic in nature \bea \delta_{Q}
\hat{\psi}_{i}^{~\hat{\alpha}}~&=&~\epsilon^{I}(R_{I})_{j}^{~\hat{\alpha}}\hat{G}_{i}^{~j}~,\nonumber\\
\delta_{Q}\hat{G}_{i}^{~j}~&=&~i\epsilon^{I}(L_{I})^{~j}_{\hat{\alpha}}\partial_{\tau}\hat{\psi}_{i}^{~\hat{\alpha}}~,\label{mm}
\eea and is inserted through the action \be S_{M}~=~\int
d\tau[i\hat{\psi}_{\hat{\alpha}}^{~i}\partial_{\tau}\hat{\psi}_{i}^{~\hat{\alpha}}+\hat{G}_{j}^{~i}\hat{G}_{i}^{~j}+M\hat{G}_{i}^{~i}]~.\label{mmact}\ee
Here the bosonic auxiliary trace ${\cal G}_{i}^{~i}$ plays a
different role with respect to the other off-trace component
because it is responsible, by its equation of motion, for setting
the mass equal to $M$. It can be easily recognized the resemblance
between the Sherk-Shwartz method \cite{SS} and the above way to
proceed if we interpret the mass multiplet as a (D+1)-th Minkowski
momentum component without coordinate analogue. We underline that
the "mass multiplet" (\ref{mm}) is crucial if we want to insert
the mass and preserve the preexisting symmetries, as it happens
for the $\psi_{5}$ field (\ref{psi5}).

\subsubsection{First and second-order formalism for spinning particle coupled to minimal N-extended supergravity}

For completeness, we include the coupling of the above models to
minimal 1-dimensional supergravity. The supergravity multiplet
escape from ${\cal GR}(d,N)$ embedding because its off-shell
fields content $(e,\chi_{I})$, \bea \delta_{Q}e~&=&~-4ie^{2}\epsilon_{I}\chi_{I}~,\nonumber\\
\delta_{Q}\chi^{I}~&=&~-\partial_{\tau}\epsilon_{I}~, \eea
consists of one real boson and $N$ real fermions. General
coordinate variations are \bea
\delta_{GC}e~=~\dot{e}\xi-e\dot{\xi}~,\nonumber\\
\delta_{GC}\chi^{I}~=~\partial_{\tau}(\chi^{I}\xi)~. \eea The
gauging of supersymmetry require the introduction of connections
by means of generators ${\cal A}_{IJ}$ valued on an arbitrary Lie
algebra. Local supersymmetric variations comes from
(\ref{USPMtransf}), (\ref{transf3}) and (\ref{transf4}) by
replacing \be
\partial_{\tau}\rightarrow {\cal
D}~=~e\partial_{\tau}+e\chi^{I}Q_{I}+\frac{1}{2}w^{JK}{\cal
A}_{JK}~,\ee while the gravitino one can be written \be
\delta_{Q}\chi^{I}~=~-[\delta_{J}^{~I}\partial_{\tau}-\frac{1}{2}e^{-1}w^{KL}(f_{KL})_{J}^{~I}]\epsilon^{J}~.
\ee One can explicitly check that the local supersymmetric invariant
action for the above mentioned transformations, is \bea S~=~\int
d\tau \{&& e^{-1}[P{\cal D}_{\tau}X+\frac{1}{d}{\cal
G}_{i}^{~j}{\cal F}_{j}^{~i}-i\bar{\psi}^{I}{\cal D}_{\tau}\psi_{I}+
\frac{i}{d}{\cal X}_{i}^{~\hat{\alpha}}\Lambda^{~i}_{\hat{\alpha}}\nonumber\\
&&-\frac{1}{2}P^{2}-\frac{1}{2d}{\cal G}_{i}^{~j}{\cal
G}_{j}^{~i}-\frac{i}{2}\bar{\psi}^{I}{\cal
D}_{\tau}\bar{\psi}_{I}-\frac{i}{2d}{\cal
X}_{i}^{~\hat{\alpha}}{\cal D}_{\tau}{\cal
X}_{\hat{\alpha}}^{~i}]\nonumber\\
&&-i\chi_{I}[\bar{\psi}_{I}P+\frac{1}{d}(f_{IJ})_{j}^{~i}{\cal
G}_{i}^{~j}\bar{\psi}_{J}+\frac{1}{d}(L_{I})_{\hat{\alpha}}^{~i}{\cal
G}_{i}^{~j}{\cal X}_{j}^{~\hat{\alpha}} \nonumber\\
&&\psi_{I}P+\frac{1}{d}(f_{IJ})_{j}^{~i}{\cal
G}_{i}^{~j}{\psi}_{J}-\frac{1}{d}(L_{I})_{\hat{\alpha}}^{~i}{\cal
F}_{i}^{~j}{\cal X}_{j}^{~\hat{\alpha}} ]~~\}~,\nonumber\\
\label{Nextsugra}\eea providing the first-order massless model for
spinning particle minimally coupled to N-extended supergravity on
the worldline. Equation of motion associated to $P$ field reads
\be P~=~{\cal
D}_{\tau}X-i\chi_{I}\bar{\psi}_{I}-i\chi_{I}\psi_{I}~,\ee that,
substituted in (\ref{Nextsugra}), give us the second-order
formulation \bea S~=~\int d\tau \{&& \frac{1}{2}{\cal
D}_{\tau}X{\cal D}_{\tau}X+\frac{1}{d}{\cal G}_{i}^{~j}{\cal
F}_{j}^{~i}-i\bar{\psi}^{I}{\cal D}_{\tau}\psi_{I}+
\frac{i}{d}{\cal X}_{i}^{~\hat{\alpha}}\Lambda^{~i}_{\hat{\alpha}}\nonumber\\
&&-\frac{1}{2d}{\cal G}_{i}^{~j}{\cal
G}_{j}^{~i}-\frac{i}{2}\bar{\psi}^{I}{\cal
D}_{\tau}\bar{\psi}_{I}-\frac{i}{2d}{\cal
X}_{i}^{~\hat{\alpha}}{\cal D}_{\tau}{\cal
X}_{\hat{\alpha}}^{~i}\nonumber\\
&&-\frac{i}{d}\chi_{I}[(f_{IJ})_{j}^{~i}{\cal
G}_{i}^{~j}\bar{\psi}_{J}+(L_{I})_{\hat{\alpha}}^{~i}{\cal
G}_{i}^{~j}{\cal X}_{j}^{~\hat{\alpha}}\nonumber\\
&&(f_{IJ})_{j}^{~i}{\cal
G}_{i}^{~j}{\psi}_{J}-(L_{I})_{\hat{\alpha}}^{~i}{\cal
F}_{i}^{~j}{\cal X}_{j}^{~\hat{\alpha}}+\chi\chi terms ]~~\}~.\eea
Finally, the massive theory is obtained following the ideas of the
previous paragraph. The massive supermultiplet (\ref{mm}) is
coupled to supergravity supermultiplet by the extension of
(\ref{mmact}) to the local supersymmetric case \bea &&S_{M}~=~\int
d\tau\{ie^{-1}\hat{\psi}_{\hat{\alpha}}^{~i}\partial_{\tau}\hat{\psi}_{i}^{~\hat{\alpha}}
+e^{-1}\hat{G}_{j}^{~i}\hat{G}_{i}^{~j}+e^{-1}M\hat{G}_{i}^{~i}
\nonumber\\
&&~~~~~~~~~~~~~~~~~~~~+i\chi^{I}(L_{I})_{\hat{\alpha}}^{~i}\hat{\psi}^{\hat{\alpha}}_{~j}\hat{\cal
G}_{i}^{~j}-iM\chi^{I}(R_{I})_{i}^{~\hat{\alpha}}\hat{\psi}_{\hat{\alpha}}^{~i}\}~,
\eea that is exactly what we need to add to the action
(\ref{Nextsugra}) to achieve a completely off-shell massive
first-order description. We close this review by noting that
importance issues regarding zero-modes of the models discussed
above have yet to be resolved.  So we do not regard this as a
completed subject yet.

\subsection{ $N=8$ unusual representations}

There exists also some ``unusual'' representations in this
approach to 1D supersymmetric quantum mechanics.  As an
illustration of these, the discussion will now treat such a case
for ${\cal {GR}}(8,8)$. It may be verified that a suitable
representation is provided by the 8 $\times$ 8 matrices \be
\begin{array}{ccccccccccccccccccc}
~{\rm L}_{1} \,&=~ i  \sigma^3 & \otimes & \sigma^2 & \otimes &
\sigma^1 &{}& {} &,~~& {\rm L}_{5} \,&=~ i \sigma^1 & \otimes &
\sigma^1  & \otimes &   \sigma^2
&{}& {} ~~, \\
~{\rm L}_{2} \,&=~ i\sigma^3 & \otimes &   \sigma^2 & \otimes &
\sigma^3 &{}& {} &,~~& {\rm L}_{6} \,&=~  i \sigma^1 & \otimes &
\sigma^3  & \otimes & \sigma^2
&{}& {} ~~, \\
~{\rm L}_{3} \,&=~  i\,   \sigma^3 & \otimes &  {\rm I}_2   &
\otimes & \sigma^2 &{}& {} &,~~& {\rm L}_{7} \,&=~ i \sigma^1 &
\otimes &   \sigma^2  & \otimes &   {\rm I}_2
&{}& {}  ~~, \\
~{\rm L}_{4} \,&=~ - \, i \sigma^2 & \otimes & {\rm I}_2   &
\otimes &  {\rm I}_2   &{}& {} &, ~~& {\rm L}_{8} \,&=~ \,  {\rm
I}_2\ & \otimes & {\rm I}_2 & \otimes & {\rm I}_2 &{}&  {} ~~.
\end{array}
\label{ultra1}\ee

An octet of scalar fields $A_{I}$ and spinor fields $\Psi_{I}$ may
be introduced. The supersymmetry variation of these are defined by
\be \delta_{Q} A_{  J} ~=~ i \epsilon^{I} (L_{  J})_{  I K} \Psi_{
K} ~~~,~~~ \delta_{Q}  \Psi_{  K} ~=~ - \, \epsilon^{I} \,(R_{
N})_{  K I}
 {\Big (}\partial_{\tau} A_{  N} {\Big )}
\label{ultra2}\ee where I, J, K, etc. now take on the values 1,
2,...,8.  Proper closure of the supersymmetry algebra requires in
addition to (\ref{cond1}) also the fact that \be (R_{  N})_{  K
J}(L_{  N})_{ I M}~+~ (R_{ N})_{ K I}(L_{  N})_{  J M} ~=~
-~2\delta_{  I J}\delta_{\rm K M}~~~~, \label{ultra3}\ee which may
be verified for the representation in (\ref{ultra1}). This is the
fact that identifies the representation in (\ref{ultra2}) as being
an `unusual' representation. The representation in (\ref{ultra2})
through the action of various AD maps generates many closely
related representations.  In fact, it can be seen that for any
integer $p$ (with $0 \le p \le 8$) there exist (p, 8, 8 - $p$)
representations!
\begin{table}

\begin{center}

\begin{tabular}{|c|c| }

\hline ~$~~~p~~$~&$ ~~Degeneracy~~$
\\ \hline  \hline
$~~~8~~$~&$ 1$
\\ \hline
$~~~7~~$~&$ 2$
\\ \hline
$~~~6~~$~&$ 1$
\\ \hline
$~~~5~~$~&$ 2$
\\ \hline
$~~~4~~$~&$ 4$
\\ \hline
$~~~3~~$~&$ 2$
\\ \hline
$~~~2~~$~&$ 1$
\\ \hline
$~~~1~~$~&$ 2$
\\ \hline
$~~~0~~$~&$ 1$
\\ \hline
\end{tabular}\\[.1in]
\caption{The representation in (\ref{ultra2}) has been given the
name ``ultra-multiplets.''  The sum of the degeneracies  adds to
sixteen.}
\end{center}
\label{redx}
\end{table}

Table 5 shows that there are, for example, two distinct
supermultiplets that have seven propagating bosons. In order to
gain some insight into how this profusion of supermultiplets comes
into being, it is convenient to note that the matrices in
(\ref{ultra1}) can be arranged according to the identifications

\bea  \alpha_{\hat A} &=& \left[\begin{array}{c}
~L_1 \\
~L_2 \\
~L_3 \\ \end{array} \right]~, \quad\quad \beta_{\hat A} =
\left[\begin{array}{c}
~L_5 \\
~L_6 \\
~L_7 \\ \end{array} \right]~, \nonumber\\ \Theta &=&L_4 ~,
\quad\quad\quad ~~~~ L_8 = \delta ~, \label{ultra4} \eea where the
quantities $L_{\rm I}$ is split into triplets of matrices
$\alpha_{\hat A}$ and $\beta_{\hat A}$, as well as the single
matrix $\Theta$ and the identity matrix $\delta$.

With respect to this same decomposition the eight bosonic fields
may be written as \be
 A_{  I} ~=~ \{
~ {\cal P} _{\hat A} ,\, ~ {\cal A}_{\hat A} ,\, ~ {\cal A} ,\,
 ~ {\cal P} \}~.
\label{ultra5}\ee Now the two distinct cases where $p$ $=$ 2 occur
from the respective AD maps

\be
 \{
~ {\cal P} _{\hat A} ,\, ~ {\cal A}_{\hat A} ,\, ~ {\cal A} ,\,
 ~ {\cal P} \}
~~\to~~  \{ ~ {\cal P} _{\hat A} ,\, ~ {\cal A}_{\hat A} ,\, ~
\partial_{\tau}^{-1} {\cal A} ,\,
 ~ {\cal P} \}
\label{ultra6}\ee and \be  \{ ~ {\cal P} _{\hat A} ,\, ~ {\cal
A}_{\hat A} ,\, ~ {\cal A} ,\,
 ~ {\cal P} \}
 ~~\to~~  \{
~ {\cal P} _{\hat A} ,\, ~ {\cal A}_{\hat A} ,\, ~ {\cal A} ,\,
 ~ \partial_{\tau}^{-1}{\cal P} \}.
\label{ultra7}\ee

\section{Graphical supersymmetric representation technique: Adinkras}

The root labels defined in (\ref{rootlabel}) seems to be good
candidates to classify linear representation of supersymmetry.
However a more careful analysis reveals that there is not a one to
one correspondence between admissible transformations and labels.
For instance the $N=1$ scalar supermultiplet can be identified by
both $(0,0)_{+}$ and $(0,1)_{-}$ root labels.

To fully exploit the power of the developed formalism, we need to
introduce a more fundamental technique that from one side
eliminate all the ambiguities and from another side reveal new
structures. A useful way to encode all the informations contained
in a supermultiplet, is provided by a graphical formulation where
each graph is christen adinkra in honour of Asante populations of
Ghana, West Africa, accustomed to express concepts that defy usual
words, by symbols. This approach was pioneered in \cite{FG}.

Basic pictures used to represent supersymmetry are circles
(nodes), white for bosons and black for fermions component fields,
connected by arrows that are chosen in such a way to point the
higher component field which is assumed to be the one that does
not appear differentiated in the r.h.s. of transformation
properties. The general rule to follow in constructing variations
from adinkras is \be \delta_{Q}f_{i}~=~\pm
i^{b}\partial_{\tau}^{a}f_{j}~,\label{genrule}\ee where
$f_{i},~f_{j}$ are two adjacent component fields, $b=1$ ($b=0$) if
$f_{j}$ is a fermion (boson) and $a=1$ ($a=0$) if $f{j}$ is the
lower (higher) component field. The sign has to be the same for
both the nodes connected. Its relevance became clear only for
$N>1$ as will be discussed in the paragraph 4.1. In the following
we introduce a general procedure to classify root tree
supermultiplets, that works for arbitrary $N$.

\subsection{$N=1$ supermultiplets}

It is straightforward to recognize the $N=1$ scalar supermultiplet
labelled by $(0,0)_{+}$\be
\begin{array}{ccccccccccccccccc}
\delta_{Q}\phi&~=~&i\epsilon\psi~, \\
\delta_{Q}\psi&~=~&\epsilon\partial_{\tau}\phi~,
\end{array}
~~~~~~\Rightarrow~~~~~~\begin{array}{ccccccccccccccccc}
\\
\\
\includegraphics[width=0.15in]{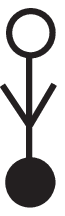}\\
\end{array} \label{scalar}
\ee and the first level dualized supermultiplet $(1,0)_{+}$ \be
\begin{array}{cccccccccccccccc}
\delta_{Q}\phi&~=~&i\epsilon\partial_{\tau}\psi~, \\
\delta_{Q}\psi&~=~&\epsilon \phi~.
\end{array}~~~~~~\Rightarrow
~~~~~~\begin{array}{ccccccccccccccccc}
\\
\\
\includegraphics[width=0.15in]{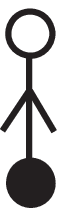}
\end{array} \label{spinor}
\ee that corresponds to the spinorial supermultiplet. The order of
the nodes is conventionally chosen to keep contact with component
fields order of the bosonic root labels (i.e. marked with a plus
sign).
Alternatively, starting from (\ref{scalar}), we can dualize the
second level, falling in the (\ref{spinor}) option. The last
possibility is to dualize both levels but again we go back to the
scalar case. Now that we have run out all the bosonic root label
possibilities, we can outline the following sequence of
congruences \bea &&(0,0)_{+}~\simeq~(1,1)_{+}~,\nonumber\\
&&(1,0)_{+}~\simeq~(0,1)_{+}~.\eea  In this framework the AD
(\ref{ad}) is seen to be implemented by a simple change in the
orientation of the arrow. We refer to this simple sequence, made
of all the inequivalent root tree supermultiplets of the bosonic
type, as the ``base sequence''.

Beside the AD we have another kind of duality, namely the Klein
flip \cite{jim4}, which corresponds to the exchanging of bosons
and fermions.  If we apply the Klein flip to the previous adinkras
it happens that we get what we call the ``mirror sequence'' \bea
\begin{array}{ccccccccccccccccc}
\\
\includegraphics[width=0.15in]{Ascalar1.eps}
\end{array} ~~~~~~\rightarrow ~~~~~~\begin{array}{ccccccccccccccccc}
\\
\includegraphics[width=0.15in]{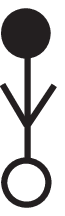}
\end{array}~~~~~~\Rightarrow ~~~~~~\begin{array}{cccccccccccccccc}
\delta_{Q}\psi&~=~&\epsilon \phi~,\\
\delta_{Q}\phi&~=~&i\epsilon\partial_{\tau}\psi~,
\end{array}\label{spinor-}\\
\begin{array}{ccccccccccccccccc}
\\
\includegraphics[width=0.15in]{Aspinor1.eps}
\end{array} ~~~~~~\rightarrow ~~~~~~\begin{array}{ccccccccccccccccc}
\\
\includegraphics[width=0.15in]{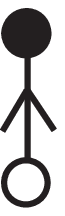}
\end{array}~~~~~~\Rightarrow ~~~~~~\begin{array}{ccccccccccccccccc}
\delta_{Q}\psi&~=~&\epsilon\partial_{\tau}\phi~,\\
\delta_{Q}\phi&~=~&i\epsilon\psi~.
\end{array}\label{scalar-}\eea

Here the KF is responsible of a changing of the supermultiplets
nature from bosonic to fermionic. Accordingly, to maintain the
order of fermionic root labels, we put a fermionic node on the
upper position. As in the base sequence, even in the mirror one,
we have congruences between root labels, precisely
$(0,0)_{-}\simeq(1,1)_{-}$ are referred to (\ref{spinor-}) while
$(1,0)_{-}\simeq(0,1)_{-}$ to (\ref{scalar-}). The power of
adinkras became manifest when we try to find which supermultiplet
of the base sequence is equivalent to the supermultiplets in the
mirror one. It is straightforward to see that up to 180 degree
rotations, only two adinkras are inequivalent \bea
(0,0)_{+}~\simeq~(1,1)_{+}~\simeq~(0,1)_{-}~\simeq~(1,0)_{-}~,\nonumber\\
(0,0)_{-}~\simeq~(1,1)_{-}~\simeq~(0,1)_{+}~\simeq~(1,0)_{+}~.\eea
All the above results about $N=1$ root tree supermultiplet, can be
reassumed in a compact way in the table
\begin{table}
\begin{center}
\includegraphics[width=1.5in]{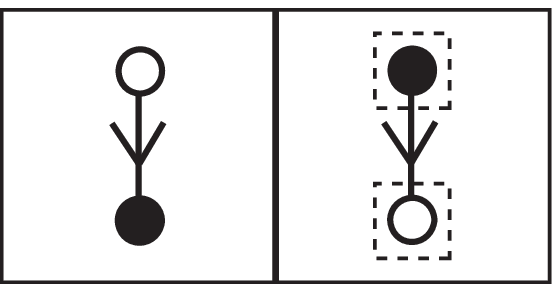}
\caption{N=1 root tree elements.}
\end{center}
\end{table}

 where boxed nodes refer to
auxiliary fields. Actually there exists a way to define auxiliary
fields by means of adinkras without appealing to the dynamics. In
the following we will denote as auxiliary all the fields whose
associated bosonic (fermionic) nodes are sink (source), namely all
the arrows point to (comes out from) the node.

\subsection{$N=2$ supermultiplets}

Even in the $N=2$ case we start from the scalar supermultiplet
(\ref{algder}) whose root label is $(0,0,0)_{+}$. Choosing the
representation \be L_{1}~=~R_{1}~=~i\sigma_{2}
,~~~~L_{2}~=~-R_{2}~=~I_{2} ~,\ee the resulting explicit
transformation
properties are \bea &&\delta_{Q}\phi_{1}~=~-i\epsilon^{1}\psi_{2}+i\epsilon^{2}\psi_{1}~,\nonumber\\
&&\delta_{Q}\phi_{2}~=~i\epsilon^{1}\psi_{1}+i\epsilon^{2}\psi_{2} ~,\nonumber\\
&&\delta_{Q}\psi_{1}~=~\epsilon^{1}\dot{\phi}_{2}+\epsilon^{2}\dot{\phi}_{1}~,\nonumber\\
&&\delta_{Q}\psi_{2}~=~-\epsilon^{1}\dot{\phi}_{1}+\epsilon^{2}\dot{\phi}_{2}
~.\label{2scalar}\eea Accordingly the adinkra associated with
(\ref{2scalar}) can be drawn as

\be
\includegraphics[width=1.5in]{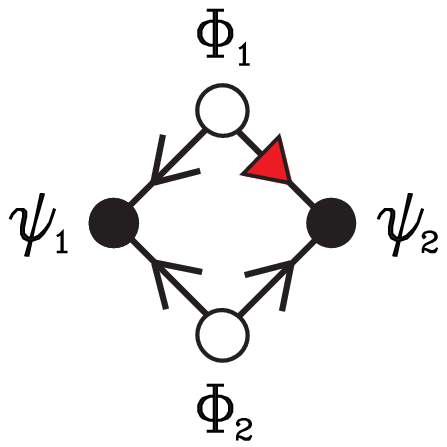}\label{Ascalar2}
\ee

The filled arrow is inserted to take into account that appears a
minus sign in the (\ref{genrule}) involving $\phi_{1}$ and
$\psi_{2}$. The $N=2$ case furnishes new features that will be
present in all higher supersymmetric extensions. One of them is
the sum rule that can be stated as follows: multiplying the signs
chosen in the (\ref{genrule}) for a closed path in the adinkra, we
should get a minus sign. Clearly the graph (\ref{Ascalar2})
satisfy this condition. Moreover it is possible to flip the sign
of a field associated to a node. The net effect on the adinkra is
a shift of the red arrow or the appearing of two more red arrows
confirming that after this kind of flip the sum rule still holds.
Another evident property is that parallel arrows correspond to the
same supersymmetry. It is easy to foresee that the $N$-extended
adinkras live in a $N$ dimensional space so that graphical
difficulties will arise for $N\geq 4$. However suitable techniques
will be developed below to treat higher dimensional cases.

The AD can be generalized to arbitrary $N$-extended cases by
saying that its application to a field is equivalent to reversing
$\underline{all}$ the arrows connected to the corresponding node.
However, as will be cleared in the next paragraphes, if we want to
move inside the root tree, we have to implement AD level by level.
This means that in (\ref{2scalar}), the AD is necessarily
implemented on both $\psi_{1}$ and $\psi_{2}$ fermionic fields.

For arbitrary  value of N, the proper way to manage the signs is
to consider the scalar supermultiplet adinkra associated to
(\ref{comptransf}), as the starting point to construct all the
other root tree supermultiplets implementing AD, Klein flip and
sign flipping of component fields. Since the scalar supermultiplet
has well defined signs by construction, the resulting adinkras
will be consistent with the underlying theory. This allows us to
forget about the red arrows and consider equivalent all the graphs
that differ from each other by a sign redefinition of a component
field. Once the problem of signs is understood, let us go back to
the classification problem. Following the line of the $N=1$ case,
from the scalar adinkra (\ref{Ascalar2}) we derive the base
sequence whose inequivalent graphs, with root labels on the right,
are \bea \begin{array}{ccccccccccccccccc}
\\
\includegraphics[width=0.7in]{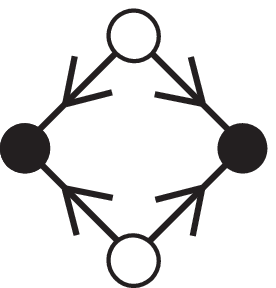}
\end{array} ~~~~~~\rightarrow ~~~~~~(0,0,0)_{+}~\simeq~(1,1,1)_{+}\label{Ascalar2+}\\
\begin{array}{ccccccccccccccccc}
\\
\includegraphics[width=0.7in]{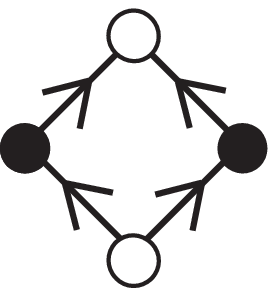}
\end{array} ~~~~~~\rightarrow
~~~~~~(0,0,1)_{+}~\simeq~(1,0,0)_{+}\label{Avector2+}\\
\begin{array}{ccccccccccccccccc}
\\
\includegraphics[width=0.7in]{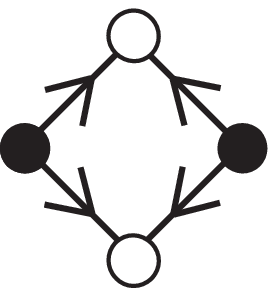}
\end{array} ~~~~~~\rightarrow ~~~~~~(0,1,0)_{+}~~~~~~~~~~~~~~~~~\label{Aspinor2+}
\eea The KF applied to the above adinkras, provides the mirror
sequence\bea \begin{array}{ccccccccccccccccc}
\\
\includegraphics[width=0.7in]{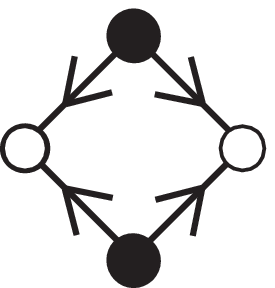}
\end{array} ~~~~~~\rightarrow ~~~~~~(0,0,0)_{-}~\simeq~(1,1,1)_{-}\label{Aspinor2-}\\
\begin{array}{ccccccccccccccccc}
\\
\includegraphics[width=0.7in]{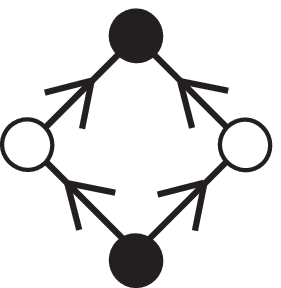}
\end{array} ~~~~~~\rightarrow
~~~~~~(0,0,1)_{-}~\simeq~(1,0,0)_{-}\label{Avector2-}\\
\begin{array}{ccccccccccccccccc}
\\
\includegraphics[width=0.7in]{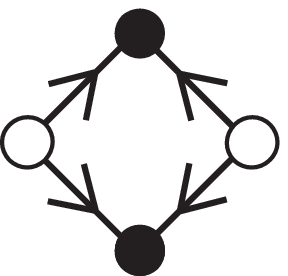}
\end{array} ~~~~~~\rightarrow ~~~~~~(0,1,0)_{-}~~~~~~~~~~~~~~~~~\label{Ascalar2-}
\eea  In order to classify the $N=2$ root tree supermultiplets,
the last step is the matching of the base sequence with the mirror
sequence to recognize topologically equivalent graphs. It turns
out that only four of them originate different dual
supermultiplets. In fact adinkra (\ref{Ascalar2+}) is nothing but
adinkra (\ref{Ascalar2-}) rotated of 90 degrees. The same relation
holds between the adinkras (\ref{Aspinor2+}) and
(\ref{Aspinor2-}). To make the remaining adinkra relationships
clear, we can arrange them in the following way \bea
\begin{array}{ccccccccccccccccc}
\\
\includegraphics[width=0.7in]{Ascalar2+.eps}
\end{array}~~~~~~~~~~\overset{KF}{\leftrightarrow}~~~~~~~~~~\begin{array}{ccccccccccccccccc}
\\
\includegraphics[width=0.7in]{Aspinor2-.eps}
\end{array}\nonumber\\
\nonumber\\
AD~\updownarrow~~~~~~~~~~~~~~~~~~~~~~~~~~~~~~~~~~~~~~~\updownarrow~AD
\nonumber\\
\begin{array}{ccccccccccccccccc}
\\
\includegraphics[width=0.7in]{Avector2+.eps}
\end{array}~~~~~~~~~~\overset{KF}{\leftrightarrow}~~~~~~~~~~\begin{array}{ccccccccccccccccc}
\\
\includegraphics[width=0.7in]{Avector2-.eps}
\end{array}\label{scheme2}
\eea so that left column is connected by the klein flip to the
right column while the upper row is the automorphic dual of the
lower one.

\subsection{Adinkras folding}
As observed in the previous paragraph, adinkras associated to
$N\geq 3$ extended supermultiplets may became problematic to draw
and consequently to classify. Fortunately there exists a very
simple way to reduce the dimensionality of the graphs preserving
the topological structure memory. The process consists in moving
the nodes and arrows into each other in a proper way. In doing
this two basic rules have to be satisfied
\begin{enumerate}
\item only nodes of the same type can be overlapped, \item we can
make arrows lay upon each other only if they are oriented in the
same way.
\end{enumerate}
In the first rule, when we talk about nodes of the same type, we
refer not only to the bosonic or fermionic nature but even to
physical or auxiliary dynamical behavior. In order to clarify how
this process works, let us graphically examine the simplest
example by folding the adinkra (\ref{Ascalar2+}) \be
\includegraphics[width=4in]{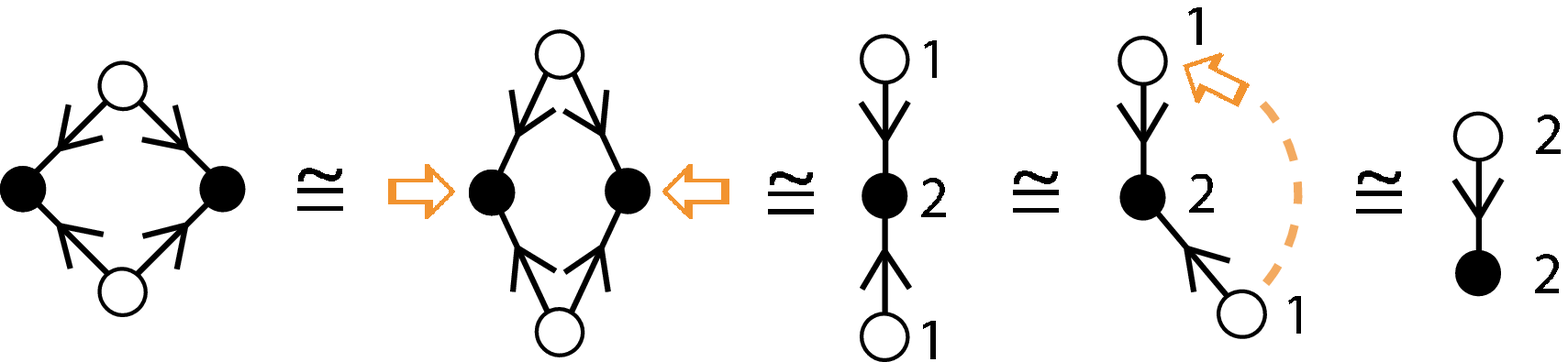}
\ee On the left of each node it is reported its multiplicity. Thus
from a 2-dim adinkra we end up to a 1-dim one, increasing the
multiplicity of the nodes. We emphasize that in the example above,
we have a sequence of two different folding. After the first one
we are left with a partially folded adinkra while in the end we
obtain a fully folded one. It is important for the following
developments, to have in mind that we have various levels of
folding for the same adinkra. A remarkable property of the root
tree elements is that they can be always folded into a linear
chain. Applying this technique to the arrangement scheme
(\ref{scheme2}), the $N=2$ root tree adinkras can be organized as
follows
\begin{table}
\begin{center}
\includegraphics[width=1.5in]{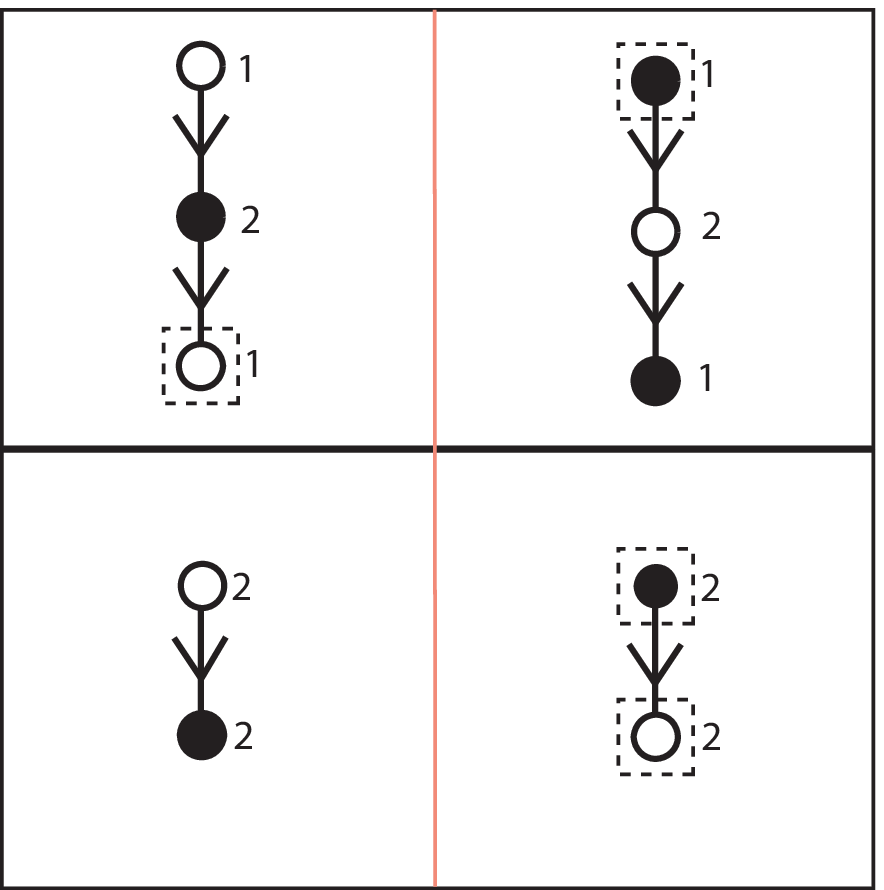}
\caption{Fully folded N=2 root tree elements.}
\end{center}
\end{table}

\subsection{Escheric supermultiplets}

In this paragraph we want to give some hints about how to describe
supermultiplets that are not in the root tree. We anticipated that
implementing AD on singular nodes may bring us outside the root
tree sequence. Let us examine this aspect in some detail. Starting
from variations (\ref{2scalar}) we dualize \be
\phi_{1}~=~\partial_{\tau}^{-1}A~, \ee to obtain \bea
&&\delta_{Q}A~=~
-i\epsilon^{1}\dot\psi_{2}+i\epsilon^{2}\dot\psi_{1}~,\nonumber\\
&&\delta_{Q}\phi_{2}~=~i\epsilon^{1}\psi_{1}+i\epsilon^{2}\psi_{2} ~,\nonumber\\
&&\delta_{Q}\psi_{1}~=~\epsilon^{1}\dot{\phi}_{2}+\epsilon^{2}A~,\nonumber\\
&&\delta_{Q}\psi_{2}~=~-\epsilon^{1}A+\epsilon^{2}\dot{\phi}_{2}
~, \eea associated to adinkra (\ref{Avector2+}). Then let the AD
map act on the left fermionic node \be
\psi_{2}~=~i\partial_{\tau}\xi~, \ee to end into \bea
&&\delta_{Q}A ~=~
-i\epsilon^{1}\ddot\xi+i\epsilon^{2}\dot\psi_{1}~,\nonumber\\
&&\delta_{Q}\phi_{2}~=~i\epsilon^{1}\psi_{1}-\epsilon^{2}\ddot\xi ~,\nonumber\\
&&\delta_{Q}\psi_{1}~=~\epsilon^{1}\dot{\phi}_{2}+\epsilon^{2}A~,\nonumber\\
&&\delta_{Q}\xi~=~i\epsilon^{1}\int^{\tilde{t}}d\tilde{t}A-i\epsilon^{2}{\phi}_{2}
~, \eea whose corresponding adinkra symbol is \be
\includegraphics[width=0.7in]{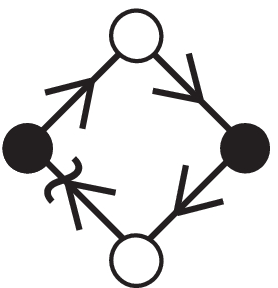}\label{EschericII} \ee where the new modified
arrow is used to describe the appearance of the antiderivative in
the r.h.s. of $\psi_{2}$ variation. Let us notice that the usual
ordering of the nodes in the adinkra (\ref{EschericII}) makes no
sense because each node is upper than the previous and lower then
the next one. This situation was one of the main theme of some
drawings of the graphic artist Maurits Cornelis Escher (see, for
instance, the lithograph ``Ascending and Descending''). For this
reason we will refer to these kind of supermultiplets as escheric.
One of the main feature of adinkra (\ref{EschericII}) is that it
can not be folded into a lower dimensional graph. This force us to
introduce a new important concept which is the rank of an adinkra,
namely the dimensions spanned by the fully folded graph diminished
by one. The case (\ref{EschericII}) provides an $N=2$ example of a
rank one adinkra while the root trees are always composed of rank
zero adinkras.

\vspace{0.5cm}

A similar result can be found even in the $N=1$ case. It is
possible to go outside the root tree enforcing the duality \be
\phi\rightarrow\partial_{\tau}^{-2}\tilde \phi~,\ee on
transformation properties (\ref{scalar}), in order to obtain \bea
\delta_{Q}\tilde{\phi}&~=~&i\epsilon\partial_{\tau}^{2}\psi~,\nonumber\\
\delta_{Q}\psi&~=~&\epsilon
\int^{\tilde{t}}d\tilde{t}~\tilde{\phi}(\tilde{t})~, \eea and the
new $N=1$ escheric symbol \be
\includegraphics[width=0.15in]{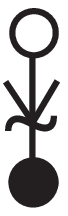} \ee
with equivalent root labels
$(2,0)_{+}\cong(0,2)_{+}\cong(1,0)_{-}\cong(0,1)_{-}$.

The integral in the r.h.s. of the above transformation properties
assume a particularly interesting meaning whenever the integrated
boson lives in a compact manifold. If this is the case then the
integral term counts the number of wrappings of the considered
bosonic field.

\vspace{0.5cm}

The above discussion about escheric supermultiplet is somehow
linked via AD maps to supersymmetric multiplets presented so far.
However, the contact with the previous approach is completely lost
by considering the adinkra \be
\includegraphics[width=0.7in]{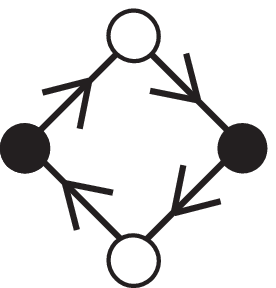}\label{EschericI}\ee
and associating to this graph the variations exploiting the
general method of equation (\ref{genrule}). It turns out that
transformation properties referred to (\ref{EschericI}) are \bea
&&\delta_{Q}\phi_{1}~=~-i\epsilon^{1}\psi_{2}+i\epsilon^{2}\dot{\psi}_{1}~,\nonumber\\
&&\delta_{Q}\phi_{2}~=~i\epsilon^{1}{\psi}_{1}+i\epsilon^{2}\dot{\psi}_{2} ~,\nonumber\\
&&\delta_{Q}\psi_{1}~=~\epsilon^{1}\dot{\phi}_{2}+\epsilon^{2}{\phi}_{1}~,\nonumber\\
&&\delta_{Q}\psi_{2}~=~-\epsilon^{1}\dot{\phi}_{1}+\epsilon^{2}{\phi}_{2}
~,\label{EschericItransf} \eea where each field is associated to
each site in the same way of adinkra (\ref{Ascalar2}) and a proper
minus sign has been inserted in order to accomplish the sum rule.
The first thing to figure out in order to understand what kind of
supersymmetric properties are hidden under this new
supermultiplet, is the commutator between two variations. It is
straightforward to prove that after the fields and parameters
complexification \bea
\phi~=~\phi_{1}+i\phi_{2}~,~~\psi~=~\psi_{1}+i\psi_{2}~,
~~\epsilon~=~\epsilon^{1}+i\epsilon^{2}~,\eea we have\footnote{the
lower indices of the supersymmetry parameters are referred to
different supersymmetries while the upper one are associated to
the two real charges of the same supesymmetry.} \be
\left[~\delta_{Q}(\epsilon_{1}),\delta_{Q}(\epsilon_{2})~\right]~=~
-2i\bar{\epsilon}_{[1}\epsilon_{2]}\partial_{\tau}
-\frac{i}{2}(\epsilon_{1}\epsilon_{2}-\bar{\epsilon}_{1}\bar{\epsilon}_{2})~\delta_{Y}\ee
where $\delta_{Y}$ acts on the fields in the following way \bea
\delta_{Y}(\phi,\psi)~&=&~(\partial_{\tau}^{2}-1)(\phi,\psi)~,\nonumber\\
\delta_{Y}(\bar{\phi},\bar{\psi})~&=&~-(\partial_{\tau}^{2}-1)(\bar{\phi},\bar{\psi})~.
\eea Expressing the variations in terms of supersymmetric charges
$Q,~\bar{Q}$ and central charge $Y$ \bea
\delta_{Q}(\epsilon)~&=&~\epsilon
Q+\bar{\epsilon}\bar{Q}\nonumber\\
\delta_{Y}~&=&~\frac{Y}{4}~,\eea we obtain the central extended
algebra \bea \{Q,~\bar{Q}\}~=~H~,~~Q^{2}~=~iY~,~~[H,~Y]~=~0~,\eea
where $Y$ plays the role of a purely imaginary central charge.
Although real central extension of similar algebras has been
studied \cite{FS}, the purely imaginary case still lacks of a
completely clear interpretation.

The escheric adinkras make clear how this graphical approach can
offer the possibility to describe theories that lie outside the
formalism developed in the previous chapters and eventually can
make arise to new non trivial features.

\subsection{Through higher $N$}

In principle, the techniques of the previous paragraphs are
suitable even for $N\geq 3$. Thus, for $N=3$ case, the scalar
supermultiplet fields transformations can be written down using
the representation for ${\cal GR}(4,3)$ given by \bea
&&L_{1}~=~R_{1}~=~i\sigma_{1}\otimes\sigma_{2}~=~\begin{pmatrix}
  _{0} & _{0} & _{0} & _{1} \\
  _{0} & _{0} & _{1} & _{0} \\
  _{0} & _{-1} & _{0} & _{0} \\
  _{-1} & _{0} & _{0} & _{0}
\end{pmatrix}~,\nonumber\\
&&L_{2}~=~R_{2}~=~i\sigma_{2}\otimes I_{2}~=~\begin{pmatrix}
  _{0} & _{1} & _{0} & _{0} \\
  _{-1} & _{0} & _{0} & _{0} \\
  _{0} & _{0} & _{0} & _{1} \\
  _{0} & _{0} & _{-1} & _{0}
\end{pmatrix}~,\nonumber\\
&&L_{3}~=~R_{3}~=~-i\sigma_{3}\otimes\sigma_{2}~=~\begin{pmatrix}
  _{0} & _{0} & _{-1} & _{0} \\
  _{0} & _{0} & _{0} & _{1} \\
  _{1} & _{0} & _{0} & _{0} \\
  _{0} & _{-1} & _{0} & _{0}
\end{pmatrix}~,\label{N=3rep}
\eea so that explicitly we have \bea
&&\delta\phi_{1}~=~-i\epsilon^{1}\psi_{4}-
i\epsilon^{2}\psi_{2}+i\epsilon^{3}\psi_{3}~,\nonumber\\
&&\delta\phi_{2}~=~-i\epsilon^{1}\psi_{3}+
i\epsilon^{2}\psi_{1}-i\epsilon^{3}\psi_{4}~,\nonumber\\
&&\delta\phi_{3}~=~i\epsilon^{1}\psi_{2}-
i\epsilon^{2}\psi_{4}-i\epsilon^{3}\psi_{1}~,\nonumber\\
&&\delta\phi_{4}~=~i\epsilon^{1}\psi_{1}+
i\epsilon^{2}\psi_{3}+i\epsilon^{3}\psi_{2}~,\nonumber\\
&&\delta\psi_{1}~=~\epsilon^{1}\dot{\phi}_{4}+
\epsilon^{2}\dot{\phi}_{2}-\epsilon^{3}\dot{\phi}_{3}~,\nonumber\\
&&\delta\psi_{2}~=~\epsilon^{1}\dot{\phi}_{3}-
\epsilon^{2}\dot{\phi}_{1}+\epsilon^{3}\dot{\phi}_{4}~,\nonumber\\
&&\delta\psi_{3}~=~-\epsilon^{1}\dot{\phi}_{2}+
\epsilon^{2}\dot{\phi}_{4}+\epsilon^{3}\dot{\phi}_{1}~,\nonumber\\
&&\delta\psi_{4}~=~-\epsilon^{1}\dot{\phi}_{1}-
\epsilon^{2}\dot{\phi}_{3}-\epsilon^{3}\dot{\phi}_{2}~, \eea that,
translated in terms of graph, are equivalent to \be
\includegraphics[width=1.2in]{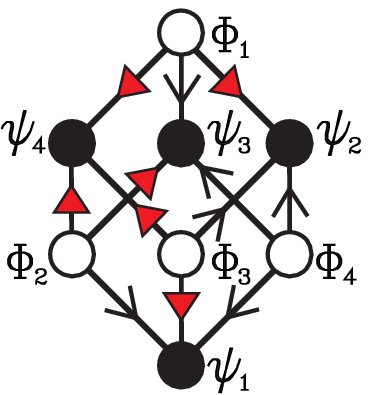}\label{Ascalar3+}\ee Next we dualize
via Klein flip to obtain \be
\includegraphics[width=0.9in]{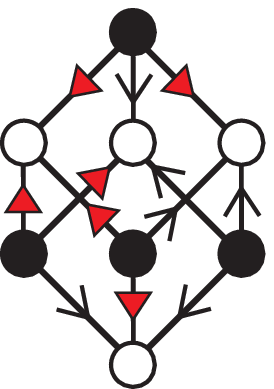}\ee that, together
with (\ref{Ascalar3+}), are respectively the starting point to
construct all the elements of the base and mirror sequences using
all possible levelwise AD. Alternatevely one can derive all the
base sequence from (\ref{Ascalar3+}) and then performs a Klein
flip on each base element in order to deduce the mirror adinkras.
Finally we fold all the topologically inequivalent adinkras
organizing them into the table
\begin{table}
\begin{center}
\includegraphics[width=4in]{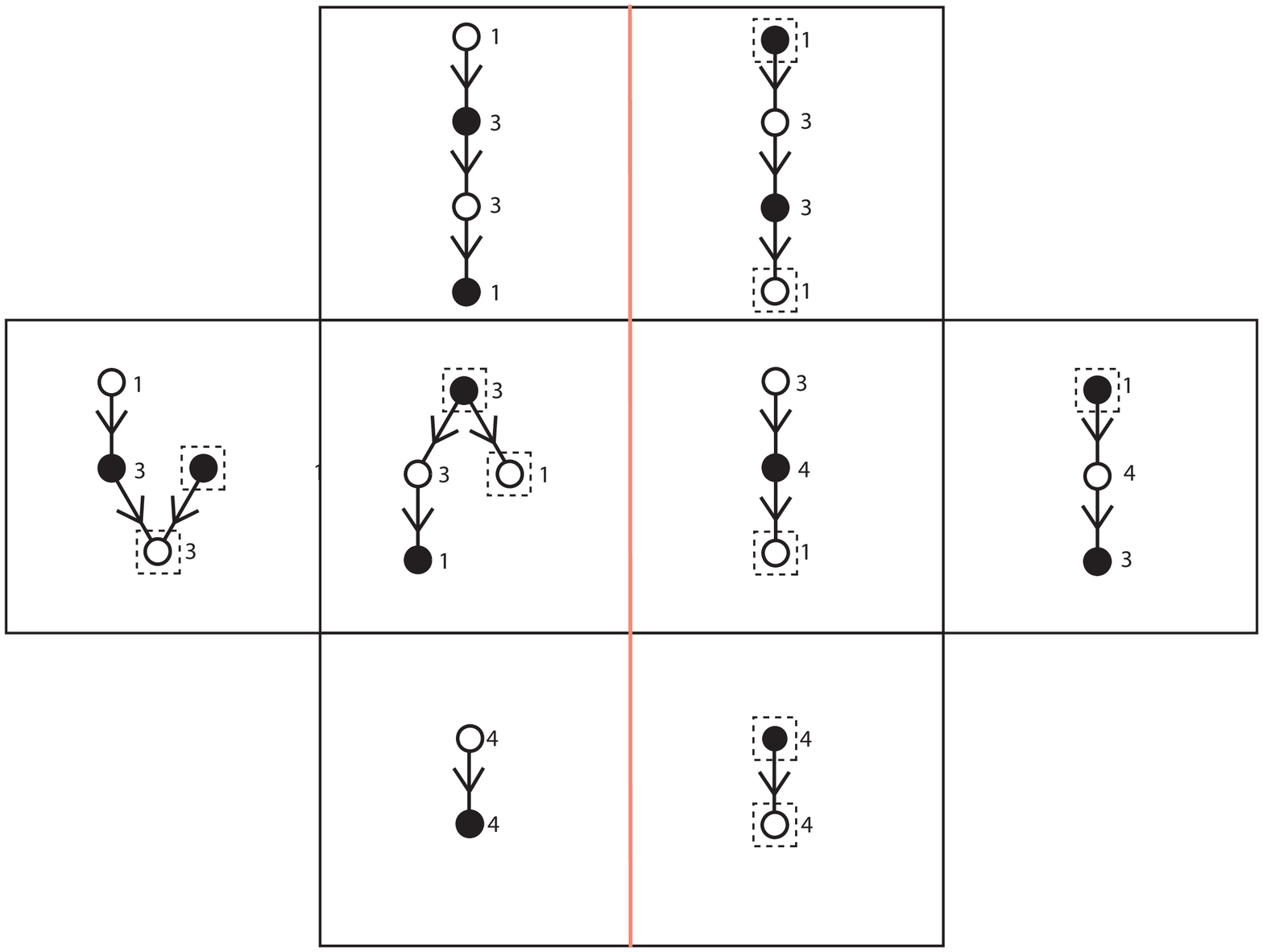} \caption{Fully folded $N=3$
root tree elements.}
\end{center}
\end{table}

As expected, all the fully folded root tree adinkras are one
dimensional. Moreover, one can verify that each closed path
without arrows within, satisfy the sum rule. The $N=3$ case
furnishes the possibility to generalize the sum rule by saying
that if $a$ is the number of arrows that are circuited by the path
then the sign of the sum rule turns out to be $(-1)^{a+1}$. It is
of some importance to notice that by completely folding the
adinkras, the levels of the nodes can be upsetted. Nevertheless if
we consider the fully unfolded one dimensional adinkras obtained
by folding the $N=3$ graphs, then we can still implement AD level
by level in order to get all the root tree. In other words the
depht of the ADs (i.e. the minimum number of dimensions reached by
the folded adinkra when the AD is applied) used to deduce the root
tree is one.

\vspace{0.5cm}

If we assume that each supersymmetry corresponds to an ortogonal
direction, as stated above, then the nodes are placed on the
vertices of a $N$ dimensional hypercube. Consequently, at $N=4$ we
find  $2^{N}|_{N=4}=16$ nodes among which eight are bosonic and
eight are fermionic. This is in contrast with the irreducible
representation dimension that is $4+4=8$ as reported in table 2.
The problem to face is how to reduce consistently the dimension of
the representation that arise from the $N=4$ adinkras in order to
obtain the irreducible representation described in the paragraph
1.3. The following two methods are effective to solve this
problem: in the first one we identify consistently the nodes to
obtain the proper transformation properties, while in the second
one we recognize irreducible sub-adinkras making rise to gauge
degrees of freedom. Let us consider the first method ( the other
one will be analyzed in the next paragraph ) constructing the
$N=4$ scalar supermultiplet as example. A possible choice of
$\cal{GR}(4,4)$ generators turns out to be composed of six
generators of the $N=3$ case (\ref{N=3rep}) plus the two
generators \be L_{4}~=~-R_{4}~=~-I_{2}\otimes I_{2}\ee It is
straightforward to figure out the following variations: \bea
&&\delta\phi_{1}~=~-i\epsilon^{1}\psi_{4}-
i\epsilon^{2}\psi_{2}+i\epsilon^{3}\psi_{3}-i\epsilon^{4}\psi_{1}~,\nonumber\\
 &&\delta\phi_{2}~=~-i\epsilon^{1}\psi_{3}+
i\epsilon^{2}\psi_{1}-i\epsilon^{3}\psi_{4}-i\epsilon^{4}\psi_{2}~,\nonumber\\
&&\delta\phi_{3}~=~i\epsilon^{1}\psi_{2}-
i\epsilon^{2}\psi_{4}-i\epsilon^{3}\psi_{1}-i\epsilon^{4}\psi_{3}~,\nonumber\\
&&\delta\phi_{4}~=~i\epsilon^{1}\psi_{1}+
i\epsilon^{2}\psi_{3}+i\epsilon^{3}\psi_{2}-i\epsilon^{4}\psi_{4}~,\nonumber\\
&&\delta\psi_{1}~=~\epsilon^{1}\dot{\phi}_{4}+
\epsilon^{2}\dot{\phi}_{2}-\epsilon^{3}\dot{\phi}_{3}-\epsilon^{4}\dot{\phi}_{1}~,\nonumber\\
 &&\delta\psi_{2}~=~\epsilon^{1}\dot{\phi}_{3}-
\epsilon^{2}\dot{\phi}_{1}+\epsilon^{3}\dot{\phi}_{4}-\epsilon^{4}\dot{\phi}_{2}~,\nonumber\\
&&\delta\psi_{3}~=~-\epsilon^{1}\dot{\phi}_{2}+
\epsilon^{2}\dot{\phi}_{4}+\epsilon^{3}\dot{\phi}_{1}-\epsilon^{4}\dot{\phi}_{3}~,\nonumber\\
&&\delta\psi_{4}~=~-\epsilon^{1}\dot{\phi}_{1}-
\epsilon^{2}\dot{\phi}_{3}-\epsilon^{3}\dot{\phi}_{2}-\epsilon^{4}\dot{\phi}_{4}~,
\eea associated to the scalar supermultiplet. Since the number of
fields are doubled by the translation into adinkras, it is
conceivable that two copies of the same adinkra could fit properly
to describe the supermultiplet. The $N=3$ scalar adinkra
(\ref{Ascalar3+}) is suitable to encode the first three
supesymmetries while the extra supersymmetry connect the nodes of
the two $N=3$ scalar adinkras copies. Graphically, the situation
can be well depicted by \be
\includegraphics[width=8cm]{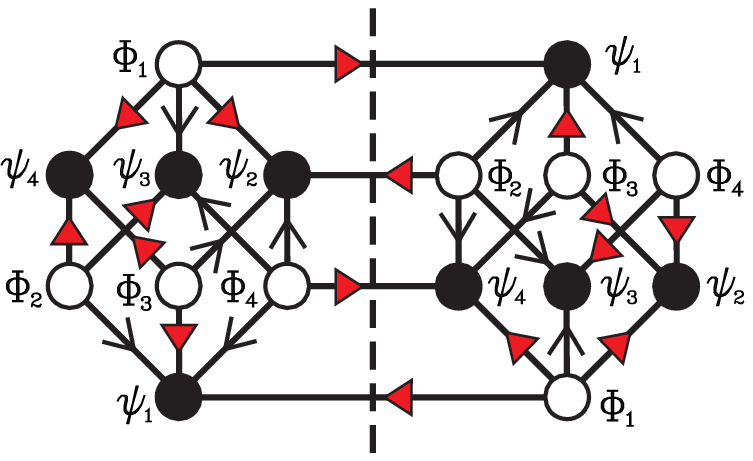}
\label{Ascalar4+}\ee where we omitted the arrows from $\phi_{3}$
to $\phi_{3}$ ( the arrows between $\phi_{2}$ and $\psi_{2}$, and
between $\phi_{2}$ and $\psi_{2}$, are not repeated in the
external nodes).We can see that the fourth supersymmetry connects
opposite nodes of the adinkra (\ref{Ascalar3+}) so that we can
render the drawning (\ref{Ascalar4+}) in the following more
compact way: \be
\includegraphics[width=4cm]{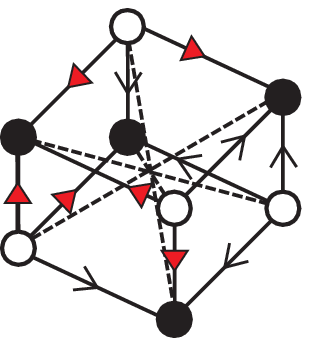}\ee where we agree
that the dashed diagonal arrows describe the same supersymmetry
even if they are not parallel. Let us notice that the second $N=3$
adinkra in the (\ref{Ascalar4+}) picture,is the mirrored copy of
the first. In fact, the dashed line in the middle represents the
mirror plane inserted to underline this property. The subtlety in
this construction is hidden in the way to connect the two $N=3$
adinkras using the fourth supersymmetry. In fact, four consinstent
choices of sign flips make rise to as many inequivalent
supermultiplets that behave to the same conjugacy class. These
four scalar supermultiplets were considered in \cite{jim4}. To
describe them it is useful to fold the adinkra (\ref{Ascalar4+})
in the following way: \be
\includegraphics[width=5cm]{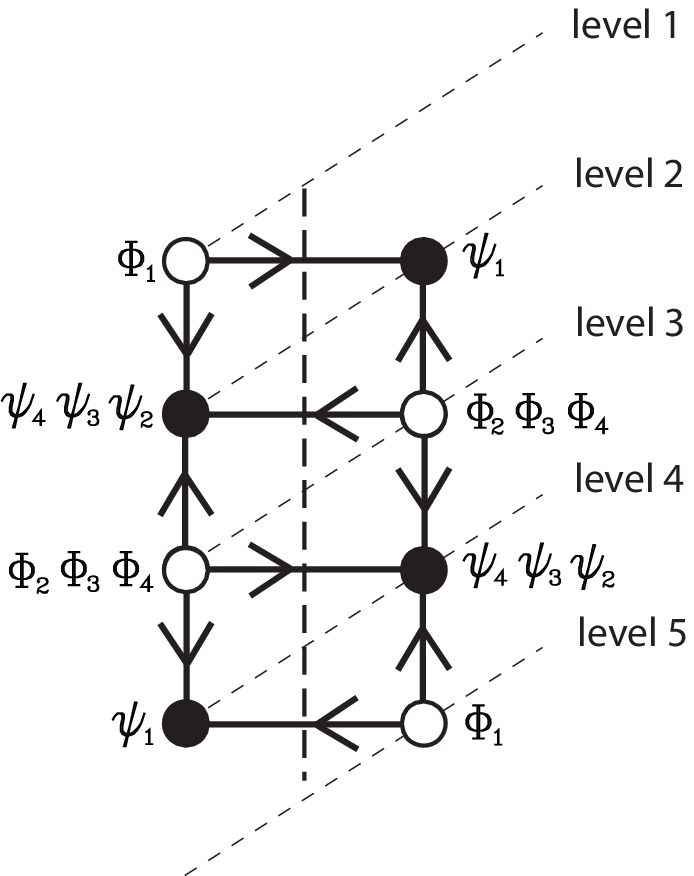}
\label{partialfolding}\ee where the diagonal dashed lines stand
for the levels of the $N=4$ supermultiplets. It is clear that the
right side with respect of the mirror plan is redundant since it
can be deduced from the left one. Therefore it is sufficient to
draw only the left side of the graph (\ref{partialfolding}) in
order to allow us to add the signs flips that identify each scalar
supermultiplet as it follows \be
\includegraphics[width=8cm]{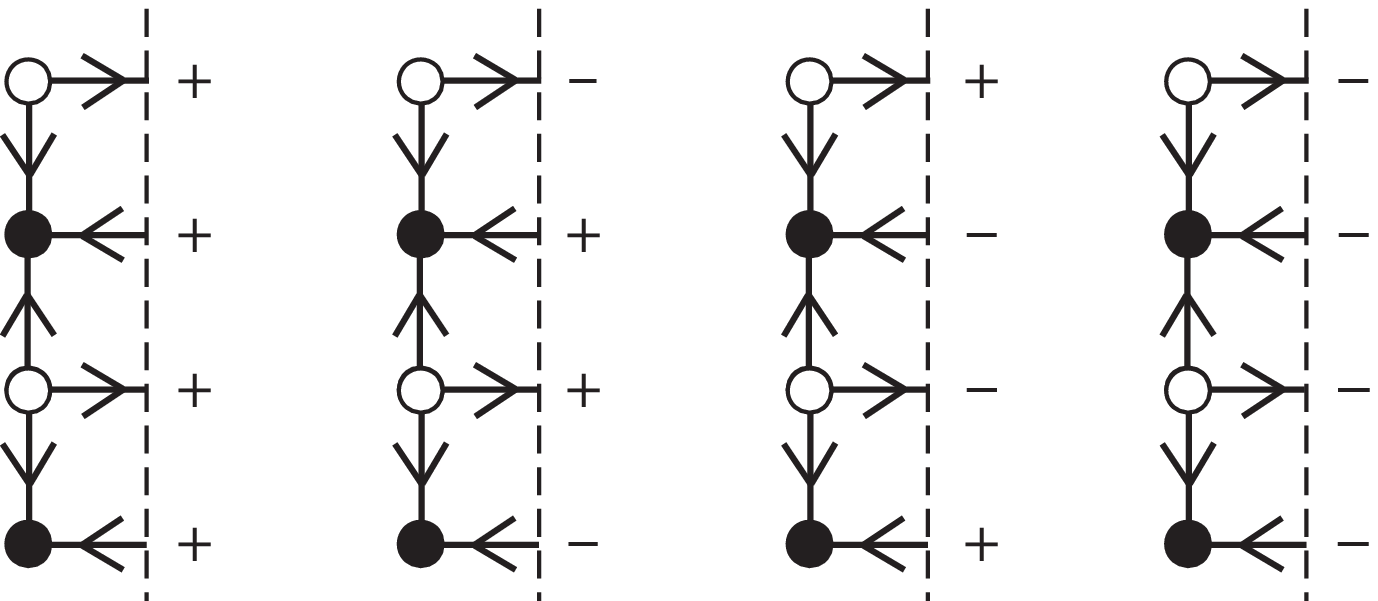}
~.\ee This quaternionic structure can be neglected if we assume
that each $N=4$ adinkra in the root tree stands for a conjugacy
class. If we adhere to this point of view, then it is a good
exercise for the reader to derive all the fully folded root tree
elements of the $N=4$ case using the techniques described so far.
The best way to proceed is to reduce the adinkra (\ref{Ascalar4+})
to its most unfolded one dimensional version which is \be
\includegraphics[width=0.5cm]{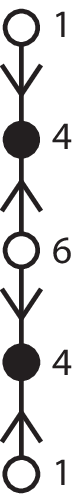}
\label{Ascalar4+1dim}\ee obtained by identifying the nodes along
the levels represented by dashed diagonal lines in the graph
(\ref{partialfolding}). We underline that the non trivial
structure of the levels is not manifest in the drawning
(\ref{Ascalar4+}) but it becames evident once we fold it into the
linear graph (\ref{Ascalar4+1dim}). One can check that the root
tree elements can be obtained dualizing along these levels and the
resulting graphs can be arranged in the table 9. The reader is
also encouraged to try to implement the AD not respecting the
suggested levels. For instance, if we consider the levels of each
N=3 sub-adinkra cube to apply AD, then it is easy to see that
escheric loops may come out.

\begin{table}
\begin{center}
\includegraphics[width=15cm]{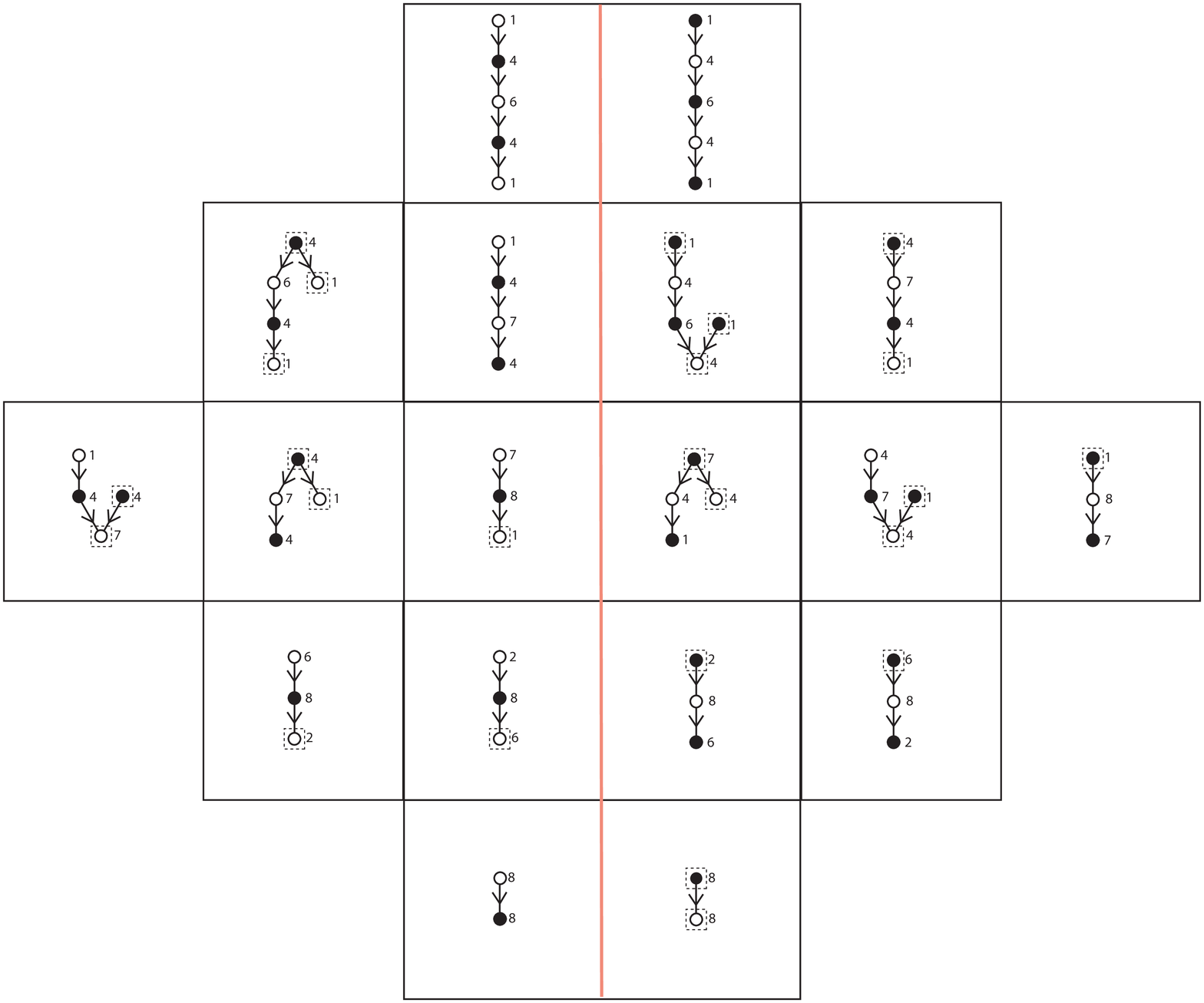}
\caption{Fully folded N=4 root tree elements.}
\end{center}
\end{table}

\subsection{Gauge invariance}

Before starting with the discussion of the gauge aspects of
adinkras, we need to describe explicitly the $N=4$ chiral
supermultiplet. To this end, let us apply a third level AD to the
adinkra (\ref{Ascalar4+}) and fold it in the following way: \be
\includegraphics[width=8.5cm]{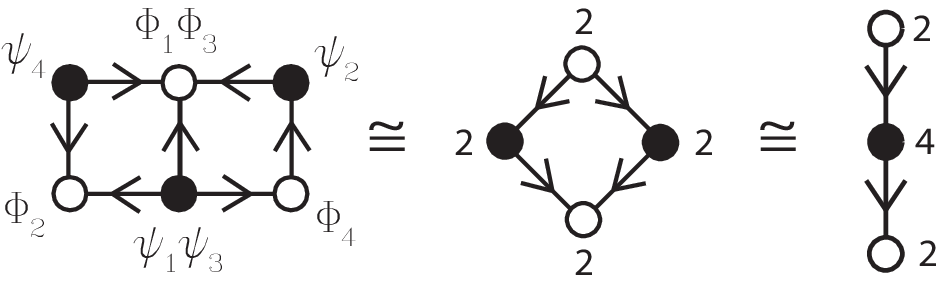}
\label{Chiral4+folding} \ee where we disregarded the second $N=3$
cube and the fourth supersymmetry arrows. The simplest root label
associated to this supermultiplet is $(00100)_{+}$ and it
corresponds to the shadow of the $N=1$ $d=4$ chiral
supermultiplet. For this reason we refer to it as the $N=4$ chiral
supermultiplet. Analogously, the shadow of $N=1$ $d=4$ vector
supermultiplet can be constructed dualizing the first, fourth and
fifth level of the scalar supermultiplet associated to the adinkra
(\ref{Ascalar4+}). By doing this we are left with \be
\includegraphics[width=5cm]{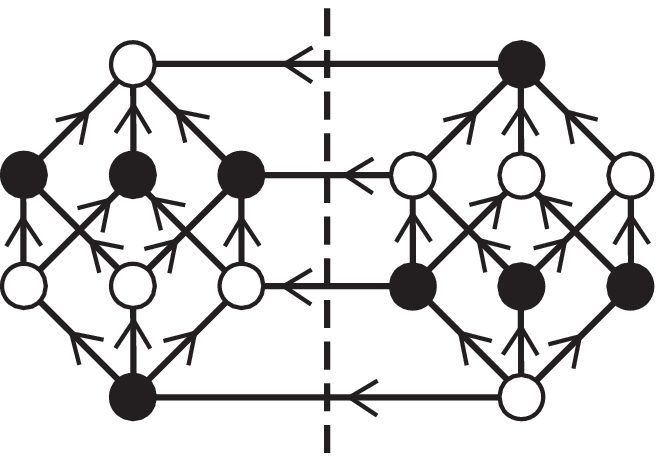}
\label{Vector4+} \ee which is foldable to the following form \be
\includegraphics[width=4cm]{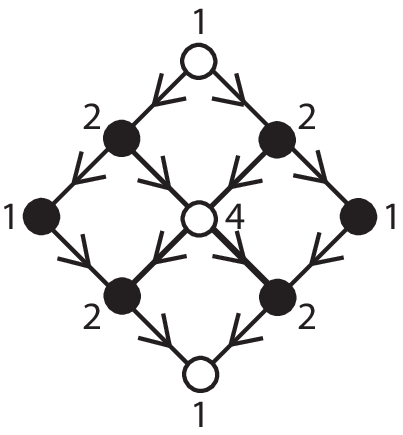} \ee In this particular structure is embedded the chiral
supermultiplet as a sub-adinkra. It is possible to remove it from
the top of the vector adinkra in order to obtain two irreducible
representations \be \includegraphics[width=12cm]{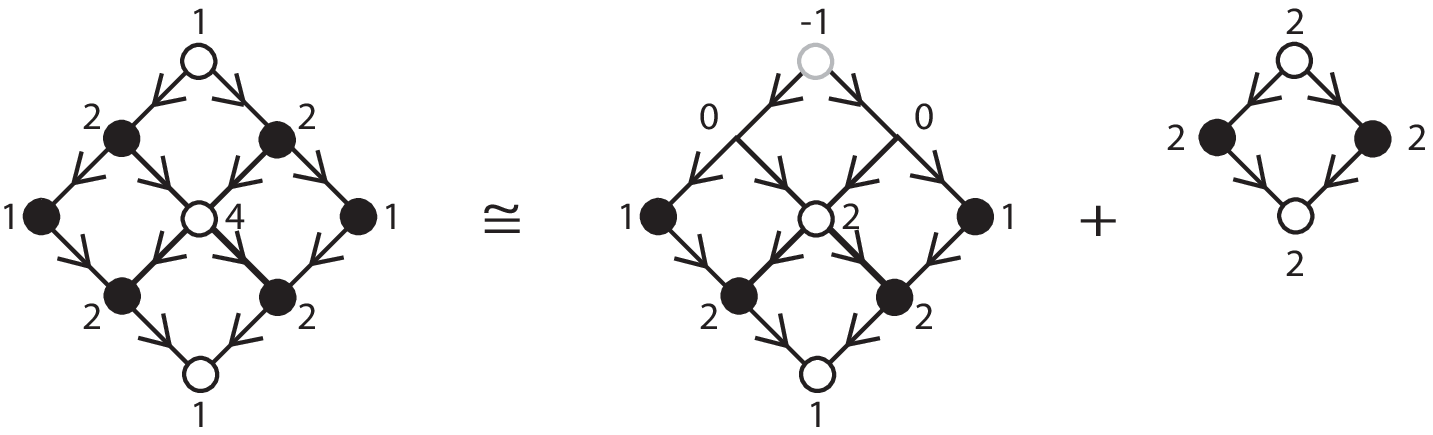} \ee
We see that a subtraction of the nodes is performed and
consequently, the topmost node of the vector adinkra assumes a
negative multiplicity. Such a node acquire the meaning of a
residual gauge degree of freedom. By moving the gauge node along
the initial structure of adinkra (\ref{Vector4+}), we fix it on
the nearest remained node, as shown in the figure \be
\includegraphics[width=9cm]{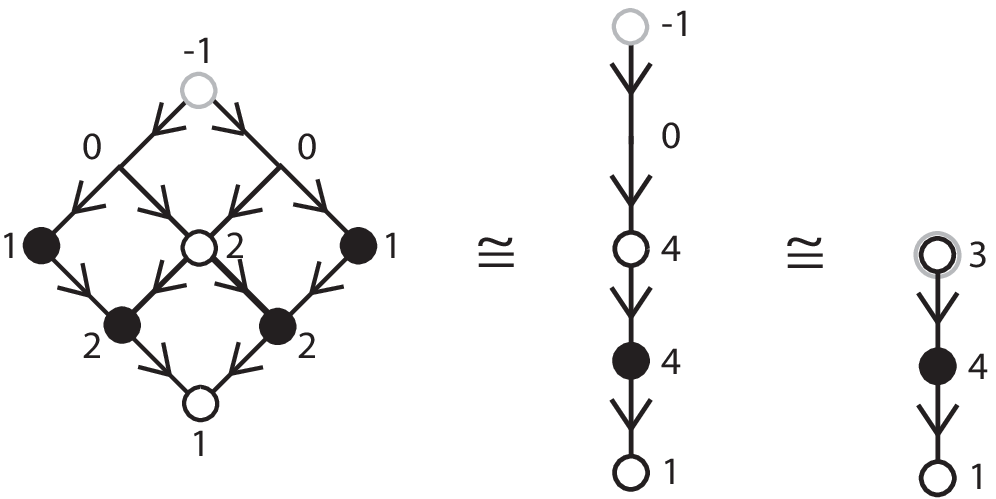} \ee This phenomenon is
nothing but the shadow of the $N=4$ $d=1$ Wess-Zumino gauge fixing
procedure. Clearly, the method described above offers an
alternative possibility to reduce the reducible supermultiplet
coming out from an adinkra simbol into two reducible
representations via the introduction of gauge degrees of freedom.

\section*{Conclusions}

By a geometrical interpretation of supersymmetric mechanics, we
reviewed a classification scheme that exploits real Clifford
algebras which are in one-to-one correspondence with the
geometrical framework of Garden Algebras. For supersymmetric
mechanics we explicitly described the link between the number of
supersymmetries and the dimension and geometry of their faithful
representations. All methods used to construct the explicit
representation of such algebras are reviewed in detail.
Particular emphasis has been dedicated to the duality relations
among different supermultiplets at fixed number of supersymmetries
using Clifford algebraic superfields. The formalism developed
turned out to be necessary, as well as effective, when applied
to the spinning particle problem, providing, quite
straightforwardly, first and second order supersymmetric actions
both in the case of global and local $N$-extended supersymmetry.
Another new application example has been provided by an $N=8$
unusual representation, suggesting how to derive many
related representation via automorphic duality.

The second part of these lectures concerned the
translation of all the results obtained so far into a simple
graphical language whose symbols are called ''Adinkras". In
particular, we encoded all properties of each supermultiplet
into an adinkra graph in order to classify and better clarify the
duality relations between supermultiplets. Using a folding
procedure to reduce the dimensions of the adinkras, we succeeded
in classifying, up to $N=4$, a large class of supermultiplets
(root trees) using linear graphs. Moreover, it has been
demonstrated that this graphical technique offers the possibility
to derive new supermultiplets through dualities, possibly with
the appearance of central charges or topological charges.

Even though the attempt to formalize a method to relate adinkras to
supermultiplets has been carried out successfully in these
lectures, many aspects still need a proper investigation on
mathematical footing. A step forward in this direction has been
presented in a recent work \cite{DFGHIL}. However, the $N\geq 4$
cases still present many unresolved classification subtleties
mainly due to the non trivial topology structure of the adinkras.
Another line of research that may be followed deals with the
implications of the duality relations between supermultiplets on
higher dimensional field theories. The oxidation procedure is a
nice tool that can be used to proceed in this way. Recently,
exploiting the automorfic duality, it has been shown \cite{BKMO}
that it is possible to relate not only the $N=4$ root tree
supermultiplets, but even the associated interacting sigma models\footnote{
Notice that in \cite{BKMO} linear and nonlinear chiral supermultiplets
were obtained by the reduction of the linear supermultiplet with
four bosonic and four fermionic degrees of freedom \cite{RussianApp,Ivanov}.}.
Anyway, if we work outside the root tree, it is still not
completely clear what kind of theories can be constructed with
such supermultiplets. Especially, it should be interesting to
better understand how to introduce central charges through
dualities. It is our belief that the techniques reviewed here
will provide new insight towards the solution of this open problem.

\section*{Acknowledgments}
E.O. would like to thank the University of Maryland and S. J.
Gates, Jr. for the warm hospitality during the development of this
work. Furthermore, E. O. would like to express his gratitude to
Lubna Rana for helpful discussions. The research of S. B. is
partially supported by the European Community's Marie Curie
Research Training Network under contract MRTN-CT-2004-005104
Forces of Universe, and by INTAS-00-00254 grant. The research of
S.J.G. is partially supported by the National Science Fundation
Grant PHY-0354401.

\end{document}